\documentclass[10pt,twocolumn,letterpaper]{article}

\usepackage[pagenumbers]{cvpr} 

\usepackage{graphicx}
\usepackage{amsmath}
\usepackage{amssymb}
\usepackage{booktabs}

\usepackage{bm}
\def\x{\bm{x}}
\def\X{\bm{X}}
\def\y{\bm{y}}
\def\Y{\bm{Y}}

\def\I{\bm{I}}
\def\T{\bm{T}}
\DeclareMathOperator{\Tr}{Tr}

\usepackage[pagebackref,breaklinks,colorlinks]{hyperref}

\usepackage[capitalize]{cleveref}
\crefname{section}{Sec.}{Secs.}
\Crefname{section}{Section}{Sections}
\Crefname{table}{Table}{Tables}
\crefname{table}{Tab.}{Tabs.}

\def\cvprPaperID{12332} 
\def\confName{CVPR}
\def\confYear{2023}

\begin{document}

\title{Sample-efficient Quantum Born Machine through Coding Rate Reduction}

\author{Pengyuan Zhai\\
Harvard University\\
Cambridge, MA, USA\\
{\tt\small pzhai@g.harvard.edu}
\and
}
\maketitle

\begin{abstract}
   The quantum circuit Born machine (QCBM) is a quantum physics inspired implicit generative model naturally suitable for learning binary images, with a potential advantage of modeling discrete distributions that are hard to simulate classically. As data samples are generated quantum-mechanically, QCBMs encompass a unique optimization landscape. However, pioneering works on QCBMs do not consider the practical scenario where only small batch sizes are allowed during training. QCBMs trained with a statistical two-sample test objective in the image space require large amounts of projective measurements to approximate the model distribution well, unpractical for large-scale quantum systems due to the exponential scaling of the probability space. QCBMs trained adversarially against a deep neural network discriminator are proof-of-concept models that face mode collapse. In this work we investigate practical learning of QCBMs. We use the information-theoretic \textit{Maximal Coding Rate Reduction} (MCR$^2$) metric as a second moment matching tool and study its effect on mode collapse in QCBMs. We compute the sampling based gradient of MCR$^2$ with respect to quantum circuit parameters with or without an explicit feature mapping. We experimentally show that matching up to the second moment alone is not sufficient for training the quantum generator, but when combined with the class probability estimation loss, MCR$^2$ is able to resist mode collapse. In addition, we show that adversarially trained neural network kernel for infinite moment matching is also effective against mode collapse. On the Bars and Stripes dataset, our proposed techniques alleviate mode collapse to a larger degree than previous QCBM training schemes, moving one step closer towards practicality and scalability.
\end{abstract}

\section{Introduction}
\label{sec:intro}
The Generative Adversarial Network (GAN) \cite{goodfellow2014generative} has been a successful generative learning framework in various scenarios \cite{cgan} \cite{salimans2016improved} \cite{infogan} \cite{bssgan} \cite{veegan} \cite{VAE-GAN}. However, GANs for discrete data have only been modeled heuristically and is an open area of research \cite{Gumbel_GAN} \cite{seqGAN}. Quantum systems, on the other hand, enjoy the naturally discrete structure and can be viewed as implicit discrete data generator, thanks to the superposition of states and measurement uncertainties \cite{Born_rule_german}. In addition, studies have proven that quantum entanglement enables quantum circuits to be more expressive than classical neural networks \cite{Bremner_PH} \cite{expressive_q}, as the correlations induced by quantum entanglement cannot be efficiently sampled classically, which is also a proposal to showcase \textit{quantum supremacy}\cite{boixo_supremacy} \cite{Lund2017QuantumSP}.

The quantum Born machine \cite{MPS_Born} \cite{benedetti-quantum-shallow} \cite{NISQ_Born} is a class of quantum-classical generative models that generates classical data bit strings by taking projective measurements on a quantum system (Born's rule \cite{Born_rule_german} \cite{born_rule}). The underlying quantum source (either real or classically simulated) can be a matrix product state \cite{MPS_Born}, a quantum circuit \cite{benedetti-quantum-shallow} \cite{NISQ_Born} \cite{dif_born}, or a dynamic evolution of a quantum system \cite{mbl_born}. In addition to learning classical data, the born machine can also reconstruct quantum states, given projective measurements as data \cite{abigail_born}. Quantum circuit Born machines (QCBMs) are described by a parameterized quantum circuit, whose parameters are updated via some classical optimization routine. Comparing with classical GANs in the continuous domain, QCBMs are naturally suitable for learning discrete images, as each measurement on the quantum system collapses the quantum state into a basis state described by bit strings (spin up or spin down). It is suggested that the QCBM can potentially produce probability distribution that is $\#$P-hard \cite{P-hard} \cite{Lund2017QuantumSP}, which cannot be simulated efficiently with classical algorithms. 

Recent studies on QCBMs focus on its gradient-based optimization \cite{dif_born} , implementability on real quantum hardwares \cite{NISQ_Born}, and expressiveness or potential supremacy \cite{Lund2017QuantumSP} \cite{expressive_q} \cite{Coyle_2020}. However, when it comes to practical learning of QCBMs (either via classical numerical simulations or on the actual quantum hardware), no previous work has sufficiently addressed training in the low batch size regime, i.e., how can we efficiently train a QCBM by only taking a very small number of projective measurements on the quantum system per optimization iteration (of updating the QCBM parameters), and make the model converge to a good learnt distribution free of mode collapse.
\\ 




Previous works have combined deep neural networks (DNNs) and QCBM models \cite{adv_Born} \cite{Situ_adv_born} \cite{born_inference}, essentially formulating a quantum-classical hybrid GAN where the generator is replaced with a QCBM, and the discriminator stays classical. These pioneering works are proofs of concepts, and do not particularly consider limited batch sizes. Mode collapse is observed in training the adversarial QCBM \cite{adv_Born}. 

In this study, we investigate QCBMs' \textit{practical} capabilities of learning binary patterns and propose sample-efficient training techniques. Namely, under small batch sizes we investigate ways to alleviate mode collapse with stable QCBM training. Based on numerical simulations on the Bars and Stripes (BAS) dataset, we confirm that the statistical two-sample test relying on the Radial Basis Function (RBF) Kernel \cite{kernel_primer} with Maximal Mean Discrepancy (MMD) loss \cite{MMD_gretton} performs poorly with small sample sizes. Meanwhile, we confirm that adversarial training with the original GAN (class probability estimation) loss \cite{goodfellow2014generative} is effective in eliminating invalid image patterns, but suffers from serious mode collapse \cite{lala2018evaluation}. We adopt moment matching as a potential cure for mode collapse. Specifically, we use the information-theoretic Maximal Coding Rate Reduction (MCR$^2$) metric \cite{yu2020learning} \cite{chan2020redu} \cite{close_loop_mcr} as a moment matching term, and interpolate it with the class probability estimation loss to study its gradual effect on the learing performance. To achieve this, we derive the gradient of MCR$^2$ with respect to quantum circuit parameters under two scenarios: i) when the feature mapping is explicit; ii) when the feature mapping is implicitly defined by a kernel. When used as a stand-alone loss function, MCR$^2$ forces diverse and subspace-like representations of data from both classical and quantum sources, but unfortunately is not sufficient for training a quantum generator. However, the MCR$^2$ term is effective when combined with the non-saturating GAN loss \cite{goodfellow2014generative}. Finally we show that an adversarially trained DNN kernel for infinite moment matching in the reproducing kernel Hilbert space (as in MMD-GAN \cite{li2017mmd} \cite{binkowski2018demystifying}) can also alleviate mode collapse in the domain of quantum born machines, which does not require searching for an additional hyper-parameter as in the  MCR$^2$-regularized non-saturating GAN loss scheme, which however comes with the cost of higher noise in the learnt model distribution.

We organize this paper as such: 
\begin{itemize}
    \item In Section \ref{sec:related_work}, we cover related work.
    \item In Section \ref{sec:Background}, we cover mathematical structures of quantum computing for readers unfamiliar with this topic. 
    \item In Section \ref{sec:alg}, we cover problem setup, model architectures and learning algorithms. We discuss likelihood-free loss functions and their underlying statistical assumptions. We also discuss the MCR$^2$ metric and derive its gradient with respect to the quantum circuit parameters.
    \item In Section \ref{sec:numerical_sims}, we showcase the effect of limited batch sizes and moment matching on the QCBM learning performance through numerical simulations.
    \item In Section \ref{sec:conclusion}, we conclude and discuss future directions.

\end{itemize}
\section{Related Work}\label{sec:related_work}
\paragraph{Generative models} Generative machine learning models (\cite{GAN-LOG}\cite{cgan}\cite{infogan}\cite{Nash-Gan} \cite{arjovsky2017wasserstein} \cite{VAE-GAN} \cite{bao2017cvae}) have enjoyed much success in recent years. Most noticeably, Generative Adversarial Networks (GANs) have been widely used for generating artificial images, text-to-image generation or image augmentation across areas of science, arts and media \cite{karras2019style} \cite{bssgan} \cite{dumoulin2016learned} \cite{text_analytics_alicia}. The goal of a generative model is to learn an set of parameters $\boldsymbol{\theta}$ that brings the model distribution $p_{\boldsymbol{\theta}}$ close to the true distribution $p^*$, based on an arbitrary choice of distance measure $d(p_{\boldsymbol{\theta}}, p^*)$.
\paragraph{Generative variational quantum circuit} Variational quantum circuits have been explored as a data generator in many studies (\cite{ref14dif}, \cite{ref15dif}, \cite{benedetti-quantum-shallow}). QCBMs (\cite{benedetti-quantum-shallow} \cite{dif_born}) exactly use the inherent probabilistic interpretation of quantum states to output discrete data samples via projective measurements on the qubits. By using a variational quantum circuit parameterized by a set of rotation angles, QCBMs can be tuned to manipulate the probability distribution, which is implicitly defined by the probabilistic output quantum states, leading to a trainable quantum circuit generator as used in studies such as \cite{benedetti-quantum-shallow}, \cite{dif_born}, \cite{Stein2020QuGANAG} \cite{adv_Born}, etc.

\paragraph{Coding Rate Reduction} Recently, \cite{yu2020learning} proposed a novel objective for learning diverse and discriminative representations of multi-class data, called ``Maximal Coding Rate Reduction'' (MCR$^2$). Models with the MCR$^2$ loss can learn underlying structures of the data, and is viable for both discriminative, auto-encoding, and unsupervised learning purposes \cite{ReduNet} \cite{close_loop_mcr} \cite{yubei_neural_cluster}. By maximizing the MCR$^2$ loss, \cite{ReduNet} maps distributions of the input data on {\em multiple} nonlinear submanifolds to multiple distinctive linear subspaces. 
\paragraph{Mode Collapse} Mode collapse happens when a GAN can only generate a few limited modes of the target data distribution \cite{GAN_survey} \cite{mode_collapse_survey}, where the elements in latent space collide to a single or very few elements in the data space \cite{mode_collapse_survey}. Hence the generator learns to produce a limited set of outcomes which the discriminator assigns high probability (of coming from the real data) to. Various GAN schemes or techniques have been proposed to alleviate mode collapse \cite{salimans2016improved} \cite{veegan_mode_collapse} \cite{che2016mode} \cite{arjovsky2017wasserstein}. Mode collapse is also prevalent in the quantum circuit born machine \cite{adv_Born} and has not been studied sufficiently.
\section{Background}\label{sec:Background}
\paragraph{Quantum state as implicit classical data distributions} The structure of quantum systems naturally supports the learning of discrete data because, for a given qubit with state $|\psi\rangle=\alpha |0\rangle+\beta |1\rangle$, $\alpha, \beta \in \mathbb{C}$, by \textbf{Born's rule} \cite{Born_rule_german}, the state collapses to either $|0\rangle$ or $|1\rangle$ with probabilities $|\alpha|^2$ and $|\beta|^2$ upon measurement. With increasing numbers of qubits, the expressive power of modeling a discrete distribution increases exponentially, as for $n$ qubits the state space grows to $|b\rangle_{1} \otimes |b\rangle_{2} \otimes \dots \otimes |b\rangle_{n}, |b\rangle_{i} \in \{|0\rangle, |1\rangle\}$. The discrete probability distribution is implicitly modeled by the $n$-qubit super-positioned quantum state $|\psi\rangle_{n} = \sum_{x=0}^{2^n-1} \alpha_x |x\rangle$, $\alpha_x\in \mathbb{C}$. $|\alpha_{x}|^2$ is the probability that upon measurement, $|\psi\rangle_{n}$ will collapse to $|x\rangle$ and we observe a classic $n$-bit string $\x$, representing (in base-two) the number between $0$ and $2^n-1$. 
Mathematically, $|x\rangle$ is equivalent to a $2^n$ element one-hot basis vector $(0,0,...,1,...0)^\top$, with a ``$1$'' only in the $x$-th position, $x=0,...,2^n-1$. Thus, for a $n$-qubit system, $\{|x\rangle\}_{x=0}^{2^n}=\{|\underbrace{000...0}_{n bits}\rangle, |000...1\rangle,...,|111..11\rangle\}=\{(\underbrace{1,0,...,0,...0}_{2^n elements})^\top, (0,1,...,0,...0)^\top,..., (0,0,...,0,...1)^\top\}$ is the \textit{standard computational basis}\cite{Nielsen_Chuang}.
\paragraph{Quantum Gates}
A quantum circuit applies quantum gates to quantum bits (or qubits). Mathematically, quantum gates are unitary matrix operations on a given quantum state vector. The set of single-qubit rotations, phase shift, and the ($2$-qubit) Controlled-NOT (CNOT) gates, can universally compute any function. The phase rotation gate $R_{\phi}$, z-axis rotation gate $R_{Z}(\theta)$, x-axis rotation gate $R_{X}(\gamma)$, and y-axis rotation gate $R_{Y}(\alpha)$ are as follows (where $\phi, \theta, \gamma, \alpha$ are rotation angles):
\begin{equation}
    \begin{gathered}
R_{\phi}=\left(\begin{array}{cc}
1 & 0 \\
0 & e^{i \phi}
\end{array}\right), \; R_{X}(\gamma)=\left(\begin{array}{cc}
\cos (\gamma / 2) & i \sin (\gamma / 2) \\
i \sin (\gamma / 2) & \cos (\gamma / 2)
\end{array}\right) \\
R_{Z}(\theta)=\left(\begin{array}{cc}
e^{i \theta / 2} & 0 \\
0 & e^{-i \theta / 2}
\end{array}\right), \\ 
R_{Y}(\alpha)=\left(\begin{array}{cc}
\cos (\alpha / 2) & \sin (\alpha / 2) \\
-\sin (\alpha / 2) & \cos (\alpha / 2)
\end{array}\right).
\end{gathered}
\end{equation}

Additionally, the CNOT gate flips the target qubit iff the the control qubit is $1$. This creates entanglement between any arbitrary pair of qubits:
$
\mathrm{CNOT}=\left(\begin{array}{llll}
1 & 0 & 0 & 0 \\
0 & 1 & 0 & 0 \\
0 & 0 & 0 & 1 \\
0 & 0 & 1 & 0
\end{array}\right). 
$

\section{Learning Model and Algorithms} \label{sec:alg}

\paragraph{Learning Task} The specific generative learning task now becomes: given a dataset $\mathcal{D}=\{\x_i\}_{i=1}^M$ composed of $M$ binary-valued images, where $\x_i$ one such image represented by a bitstring of length $N$, i.e., $\x_i\in\{0,1\}^N$. We assume these images come from a target distribution $p^*(\x)$, and the goal is to optimize the QCBM's parameters $\boldsymbol{\theta}$ such that $p_{\boldsymbol{\theta}}(\x)$ approaches to $p^*(\x)$ according to some distance measure between the two distributions. 

In this case, $\boldsymbol{\theta}$ are the rotational parameters of the quantum circuit, and we optimize them via a classical-quantum optimization scheme \. At each optimization step, we repeatedly perform measurements to collect a batch of samples (bitstrings). A classical optimizer is used to minimize the loss function $\mathcal{L}(\tilde{\X}, \X)$ with respect to model parameters $\boldsymbol{\theta}$. The loss function is based on any statistical metric that measures the distance between the model and target distribution given generated samples $\tilde{\X}$ and real samples $\X$. There is a trade-off between the number of samples generated per batch and the total number of iteration steps needed to reach convergence \cite{Situ_adv_born}.
\paragraph{Multilayer Parameterized Quantum (MPQ) Circuit}
MPQ circuits, or MPQCs, are multi-layer circuits, where each layer (or block) is composed of single-bit rotation unitaries and entanglement operations through CNOTs. The same block architecture is repeated $L$ times where $L$ is the total number of layer/depth of the circuit. We use the most generic multi-layer quantum circuit as in previous studies \cite{dif_born}. We use a circuit layout similar to \cite{ibm_vqe} and \cite{dif_born}, where we alternate single qubit rotation layers and entanglement layers. Each rotation layer $l=1,...,d$ has the form $U\left(\theta_l^j\right)=R_z\left(\theta_l^{j, 1}\right) R_x\left(\theta_l^{j, 2}\right) R_z\left(\theta_l^{j, 3}\right)$, where $j$ indexes any one of the $n$ qubits, and $R_m(\theta) \equiv$ $\exp \left(\frac{-i \theta \sigma_m}{2}\right)$ with the Pauli matrix $\sigma_m$ \cite{Nielsen_Chuang}. Including the first single $y$ rotational layers, the total number of parameters in this QCBM is $(3 d+1)n$.
\paragraph{Quantum Circuit Born Machine}
With the MPQC as its backbone data generator, Quantum circuit Born machine (QCBM) \cite{benedetti-quantum-shallow}\cite{dif_born} represents classical discrete probability distribution via the Born's rule \cite{born_boltzmann}\cite{born_rule}. QCBM utilizes a MPQC to evolve the initial/input quantum state $|z\rangle=|0\rangle^{\otimes N}$ to some target state via unitary gates: $\left|\psi_{\theta}\right\rangle=U_{\theta}|z\rangle$, where $\theta$ are the parameters of the MPQ. One measures the outputs state in the computational basis to produce a classical sample (bit string) $x \sim p_{\boldsymbol{\theta}}(x)=\left|\left\langle x | \psi_{\theta}\right\rangle\right|^{2}$. Excitingly, the output probability densities of a general quantum circuit cannot be efficiently simulated by classical means, the QCBM is among the several proposals to show quantum supremacy \cite{boixo_supremacy}. Ref. \cite{dif_born} developed a differentiable learning scheme of QCBM by minimizing the maximum mean discrepancy (MMD) loss  \cite{dif_born} \cite{MMD_gretton} using a Gaussian Kernel:
\begin{equation}
    \begin{aligned}
\mathcal{L}&=\underset{x \sim p_{\theta}, y \sim p_{\theta}}{\mathbb{E}}[K(x, y)]-2 \underset{x \sim p_{\theta}, y \sim p^*}{\mathbb{E}}[K(x, y)]\\
&+\underset{x \sim p^*, y \sim p^*}{\mathbb{E}}[K(x, y)],
\end{aligned}
\end{equation}
where $p^*$ and $p_\theta$ are the data and model distributions respectively. The function $\phi(\cdot)$ maps a data sample to an infinite-dimensional reproducing kernel Herbert space \cite{kernel_space}. This setup conveniently saves one from working in the infinite-dimensional feature space by defining a Gaussian kernel function $K(x, y)=\frac{1}{c} \sum_{i=1}^{c} \exp \left(-\frac{1}{2 \sigma_{i}}|x-y|^{2}\right)$ to only account for the pair-wise distances between two samples at various scales (the \textit{kernel trick}).

\paragraph{Generative Adversarial Quantum Circuits} QCBMs described above do not involve training a neural network. The MMD loss is directly applied on the image/data space. However, due to the lack of a learning mechanism, large sample sizes are required to approximate the MMD loss \cite{MMD_gretton} at each optimization step, which soon becomes unpractical with larger quantum systems. A neural network is thus needed to learn and adapt to seen data throughout training, allowing for stochastic gradient descent \cite{sgd} with small batch sizes. The conventional adversarial Quantum Circuits \cite{adv_Born} uses the MPQC as its generator and a classical deep neural network as the discriminator. Similar to that of a classical GAN, it follows a adversarial training scheme with class probability estimation: the (classical) discriminator takes either real samples from the dataset or synthetic samples from the quantum circuit and outputs a score, $D_{\phi}(x)$ that predicts how likely the sample is from the data distribution. The Generator tries to generate synthetic samples $x$ with high scores $D_{\phi}(x)$. The quantum generator and the classical discriminator are optimized iteratively in a zero-sum minimax game:
$$
\min _{G_{\theta}} \max _{D_{\phi}} \mathbb{E}_{x \sim p^*}\left[\ln D_{\phi}(x)\right]+\mathbb{E}_{x \sim p_{\theta}}\left[\ln \left(1-D_{\phi}(x)\right)\right].
$$

This zero-sum minimax objective \cite{goodfellow2014generative} above does not perform well for classical GANs due to vanishing gradients. The non-saturating GAN loss, writing the discriminator loss and the generator loss separately, becomes:
\begin{equation}
    \begin{aligned}
\mathcal{L}_{D_{\phi}} &=-\mathbb{E}_{x \sim p^*}\left[\ln D_{\phi}(x)\right]-\mathbb{E}_{x \sim p_{\theta}}\left[\ln \left(1-D_{\phi}(x)\right)\right] \\
\mathcal{L}_{G_{\theta}} &=-\mathbb{E}_{x \sim p_{\theta}}\left[\ln D_{\phi}(x)\right].
\end{aligned}
\end{equation}

Previous adversarial QCBM are proofs of concepts \cite{adv_Born}\cite{Situ_adv_born} and do not consider the practical regime when one can only perform limited numbers of measurements on the quantum system due to hardware restrictions nor how to deal with the mode collapse problem. Because the data generation process is quantum-mechanical, it is worth investigating what loss function families are effective in particular for training QCBMs.

\paragraph{Likelihood-free Loss functions}
Both the classical and quantum generative models are \textit{implicit probabilistic models} which generate samples without an explicit likelihood function, $p_{\theta}(\x)$ \cite{learning_in_implicit}. Due to the lack of tractable calculations of $p_{\theta}(\x)$, one only has access to samples $\tilde{\x}\sim p_{\theta}(\x)$ from the model distribution during the learning process. In the scope training QCBMs, loss functions such as maximal mean discrepancy (MMD), Wasserstein distance, class probability estimation (as in the original GAN), Sinkhorn divergence, Stein divergence have been considered \cite{Coyle_2020}, and they are nonetheless related to two families: i). class-probability estimation, e.g., the original GAN loss; ii) integral probability metric, e.g., moment matching, MMD, and Wasserstein distance, etc. The MCR$^2$ loss can be used as a likelihood-free loss function, corresponding to second-moment matching.

In this work, we particularly study the effect of moment matching for improving training QCBMs under small batch sizes.

\subsection{MCR$^2$ for Second Moment Matching}
\paragraph{Maximal Coding Rate Reduction} The \textit{coding rate} \cite{rate_distortion} measures the number of bits to encode a group of vectors (assuming $\mathbf{0}$ mean) from a multi-dimensional Gaussian source \cite{coding_rate_textbook}. In the simplest case, given two groups of samples from $p_{\X}$ and $p_{\Y}$, both with $\mathbf{0}$ mean, if we treat each sample as a column vector and construct sample matrices $\X$ and $\Y$, both are $\mathbb{R}^{D \times m}$, where $D$ is the dimension of features and $m$ is the batch size, we define the MCR$^2$ (second order) ``distance'' between sample matrices $\X$ and $\Y$ as:
\begin{equation}
\label{MCR_x_space}
    \begin{split}
        &\Delta R([\X,\Y])=\frac{1}{2}\log\det(\I + \frac{d}{2m\epsilon^2}\X\X^T+\frac{d}{2m\epsilon^2}\Y\Y^T)\\
        &-\frac{1}{4}\log\det(\I + \frac{d}{m\epsilon^2}\X\X^T)-\frac{1}{4}\log\det(\I + \frac{d}{m\epsilon^2}\Y\Y^T).
    \end{split}
\end{equation}

The MCR objective is non-negative and has interesting geometric interpretations related to ball counting in information theory \cite{rate_distortion}. This metric is $0$ iff $\frac{1}{m}\X\X^T=\frac{1}{m}\Y\Y^T$, i.e., the sample covariance matrices (given the same batch sizes) are equal (second moment matching); it is maximized when $\X^\top \Y=0$, meaning that data vectors from $\X$ and $\Y$ live in two orthogonal subspaces. Since the binary images/bitstrings follow distributions with higher moment correlations than the second moment, it is not advised to directly apply Eq. \ref{MCR_x_space} to the data space. However, if one uses a deep neural network $f_\phi$ to map high-dimensional discrete data to a low-dimensional continuous space $f_\phi: X\rightarrow Z$, where $x\in \mathbb{R}^D$ and $x\in \mathbb{R}^d$, $D\gg d$, the second moment matching becomes significant in discovering meaningful structures \cite{close_loop_mcr} \cite{yubei_neural_cluster}. In addition, if we express the sample covariance matrix in terms of probability $p(\x)$ (for the case when the sample size $m\rightarrow \infty$):
\begin{equation}
    \lim_{m\rightarrow \infty} \frac{d}{m\epsilon^2}\X\X^T=\frac{d}{\epsilon^2}\sum_{\x} p_{\X}(\x)\x\x^T,
\end{equation}
the loss function becomes (where we work in the feature vector space, abbreviating the feature mapping $f_{\phi}(\cdot)$ as $\phi(\cdot)$):
\begin{equation}
\label{infinite}
    \begin{split}
        &\Delta R([p_\theta,p^*])=\frac{1}{2}\log\det\big(\I + \mathcal{C}_{\X \Y}\big)-\frac{1}{4}\log\det\big(\I + \mathcal{C}_{\X}\big)\\
        &-\frac{1}{4}\log\det\big(\I + \mathcal{C}_{\Y}\big),
    \end{split}
\end{equation}
where
\begin{equation}
\begin{split}
    & \mathcal{C}_{\X \Y} = \frac{d}{2\epsilon^2}\left(\underset{{\x} \sim p_{\boldsymbol{\theta}}}{\mathbb{E}}\phi(\x)\phi(\x)^T+\underset{{\y} \sim p^*}{\mathbb{E}}\phi(\y)\phi(\y)^T \right), \\
    &\mathcal{C}_{\X} = \frac{d}{\epsilon^2}\underset{{\x} \sim p_{\boldsymbol{\theta}}}{\mathbb{E}}\phi(\x)\phi(\x)^T, \; \mathcal{C}_{\Y} = \frac{d}{\epsilon^2}\underset{{\y} \sim p^*}{\mathbb{E}}\phi(\y)\phi(\y)^T.
\end{split}
\end{equation}
In the first part of the investigation, we use MCR$^2$ to \textbf{study the effect of second moment matching} (of feature representations) on mode collapse behaviors in QCBMs. Second moment matching has been proposed in the QCBM context \cite{benedetti-quantum-shallow}. Due to the geometric properties of MCR$^2$ \cite{yu2020learning}, it will encourage the generator to output diverse samples to match the real feature vectors up to the second moment. Specifically, we study the model performance under various interpolation ratios ($\alpha \in [0.0, 1.0]$) between the non-saturating GAN loss and the MCR$^2$ objective for the quantum generator:
\begin{equation}
\label{interpolate}
    \begin{aligned}
\mathcal{L}_{G_{\theta}} &=(1-\alpha)\cdot\big(-\underset{x \sim p_{\theta}}{\mathbb{E}}\left[\ln D_{\phi}(x)\right]\big)+\alpha \cdot \Delta R([p_\theta,p^*]).
\end{aligned}
\end{equation}

We show later that indeed only matching up to the second moment in the feature space is not sufficient. However, second moment matching coupled with class probability estimation yields good learning results.

\subsection{Adversarial Kernel MMD for Infinite Moment Matching} 
Combining the learning power of a deep neural network and the infinite moment matching capability of MMD, we investigate a deep kernel MMD as described in MMD-GAN \cite{li2017mmd} \cite{binkowski2018demystifying}. Instead of matching the MMD distance in the image space, we adversarially train a DNN kernel $f_\phi (\cdot)$ that maps a data sample $\x$ into a low dimensional representation $f_\phi (\x)\in \mathbb{R}^d$, on which we apply the MMD loss with a Gaussian Kernel:
\begin{equation}
    \begin{aligned}
&\min_\theta \max_\phi \mathcal{L} =\underset{x \sim p_{\theta}, y \sim p_{\theta}}{\mathbb{E}}[K_\phi (x, y)]-2 \underset{x \sim p_{\theta}, y \sim p^*}{\mathbb{E}}[K_\phi (x, y)]\\
&+\underset{x \sim p^*, y \sim p^*}{\mathbb{E}}[K_\phi (x, y)],\\
&K_\phi (x, y) =\frac{1}{c} \sum_{i=1}^{c} \exp \left(-\frac{1}{2 \sigma_{i}}|f_\phi(\x)-f_\phi(\y)|^{2}\right).
\end{aligned}
\end{equation}
Even though $f_\phi$ is required by theorem $1$ of \cite{li2017mmd} to be injective, previous study suggests that relaxing this constraint still leads to good learning result, which we will follow herein.

\subsection{Differential Circuit Gradients}
The MMD loss, non-saturating GAN loss, and the MCR loss all have differentiable gradients with respect to the circuit gate parameters (rotation angles). According Ref. \cite{quantum_gradient_36}, \cite{quantum_gradient_38}, the exact gradient of the probability on value $p_\theta (x)$ with respect to a quantum circuit parameter $\theta$, is
\begin{equation}
\label{gradient}
    \frac{\partial p_{\boldsymbol{\theta}}(x)}{\partial \theta_{l}^{\alpha}}=\frac{1}{2}\left(p_{\boldsymbol{\theta}^{+}}(x)-p_{\boldsymbol{\theta}^{-}}(x)\right)
\end{equation}

where $p_{\theta^\pm} (x)=\left|\left\langle x \mid \psi_{\theta}\right\rangle\right|^{2}$ is the probability of observing $x$ generated by the same circuit with shifted circuit parameter $\theta^{\pm} \equiv \theta \pm \frac{\pi}{2}$. This is an unbiased estimate of the gradient.

All three loss function that we consider in this study, namely the MMD, non-saturating GAN and the MCR losses are sampling-based. To see this, one can apply the gradient formula (\ref{gradient}) to these three objective functions with respect to a given generator parameter $\theta$ and get (as done in \cite{dif_born}): 
\begin{equation}
    \begin{aligned}
\nabla_{\theta}\mathcal{L}_{\text{MMD}} &=\underset{x \sim p_{\theta^{+}}, y \sim p_{\theta}}{\mathbb{E}}[K(x, y)]-\underset{x \sim p_{\theta^{-}}, y \sim p_{\theta}}{\mathbb{E}}[K(x, y)]\\
&-\underset{x \sim p_{\theta^{+}}, y \sim p^*}{\mathbb{E}}[K(x, y)]+\underset{x \sim p_{\theta^{-}}, y \sim p^*}{\mathbb{E}}[K(x, y)].
\end{aligned}
\end{equation}
And for the non-saturating GAN loss (as done in \cite{adv_Born} \cite{Situ_adv_born}):
\begin{equation}
\begin{split}
    \nabla_{\theta} \mathcal{L}_{G_{\theta}}=\frac{1}{2} \underset{x \sim p_{\theta^{-}}}{\mathbb{E}}\left[\ln D_{\phi}(x)\right]-\frac{1}{2} \underset{x \sim p_{\theta^{+}}}{\mathbb{E}}\left[\ln D_{\phi}(x)\right],
\end{split}
\end{equation}
then for the MCR loss (where we abuse the notation and abbreivate the explicit feature mapping (\eg via a DNN) $f_{\boldsymbol{\phi}}(\cdot)$ as $\phi(\cdot)$), we derive the following gradient (derivations in Appendix \ref{Appendix-explicit}):
\begin{equation}
    \nabla_{\theta}\mathcal{L}_{\text{MCR}}=\nabla_{\theta}\mathcal{L}_{\text{A}}-\nabla_{\theta}\mathcal{L}_{\text{B}},
\end{equation}
\begin{equation}
    \begin{split}
        &\nabla_{\theta}\mathcal{L}_{\text{A}}=\frac{1}{4}\bigg\langle \bigg(\I + \mathcal{C}_{\X \Y} \bigg)^{-1}, \; \underset{{\hat{\x}\sim p_{\theta^+}}}{\mathbb{E}} \phi(\hat{\x})\phi(\hat{\x})^\top \bigg\rangle\\
        &-\frac{1}{4} \bigg\langle \bigg( \I + \mathcal{C}_{\X \Y} \bigg) ^{-1}, \; \underset{\hat{\x}\sim p_{\theta^-}}{\mathbb{E}} \phi(\hat{\x})\phi(\hat{\x})^\top \bigg\rangle,\\
    \end{split}
\end{equation}
\begin{equation}
    \begin{split}
        &\nabla_{\theta}\mathcal{L}_{\text{B}}=\frac{1}{8} \bigg\langle \left(\I + \mathcal{C}_{\X} \right)^{-1}, \; \underset{\hat{\x}\sim p_{\theta^+}}{\mathbb{E}} \phi(\hat{\x})\phi(\hat{\x})^\top \bigg\rangle\\
        &-\frac{1}{8}\bigg\langle \left(\I + \mathcal{C}_{\X} \right)^{-1}, \; \underset{\hat{\x} \sim p_{\theta^-}}{\mathbb{E}} \phi(\hat{\x})\phi(\hat{\x})^\top \bigg\rangle,
    \end{split}
\end{equation}

where $\langle A,B \rangle$ denotes the element-wise dot product between matrices $A$ and $B$. 

For the case \textbf{when no explicit feature mappings $\phi(\cdot)$ exist}, but only the inner product in an arbitrary kernel space is defined, \ie $K_{\phi}(\x, \y)=\phi(\x)^\top \phi(\y)$, the gradient is (derivations in Appendix \ref{Appendix-kernel}):
\begin{equation}
    \nabla_{\theta}\mathcal{L}_{\text{MCR}}=\nabla_{\theta}\mathcal{L}_{\text{A}}-\nabla_{\theta}\mathcal{L}_{\text{B}},
\end{equation}

\begin{equation}
    \begin{split}
        &2\frac{\partial \mathcal{L}_A}{\partial \hat{\theta}} \approx \frac{1}{2}\bigg(\underset{\x\sim p_{\boldsymbol{\theta^+}}}{\mathbb{E}}K_{\phi}(\hat{\x}, \hat{\x})-\underset{\x \sim p_{\boldsymbol{\theta^-}}}{\mathbb{E}}K_{\phi}(\hat{\x}, \hat{\x})\bigg)-\\
        &\frac{d}{4m\epsilon^2}\Bigg[\underset{\x \sim p_{\boldsymbol{\theta^+}}}{\mathbb{E}} K_{\phi}(\hat{\x}, \T) \Big(\I+\frac{d}{2m\epsilon^2} K_{\phi}(\T,\T)\Big)^{-1} K_{\phi}(\T, \hat{\x}) \\
        &-\underset{\x \sim p_{\boldsymbol{\theta^-}}}{\mathbb{E}} K_{\phi}(\hat{\x},\T) \Big(\I+\frac{d}{2m\epsilon^2} K_{\phi}(\T,\T)\Big)^{-1} K_{\phi}(\T, \hat{\x}) \Bigg].
    \end{split}
\end{equation}

\begin{equation}
    \begin{split}
        &4\frac{\partial \mathcal{L}_B}{\partial \hat{\theta}} \approx \frac{1}{2}\bigg(\underset{\hat{\x}\sim p_{\boldsymbol{\theta^+}}}{\mathbb{E}} K_{\phi}(\hat{\x},\hat{\x}) -\underset{\hat{\x} \sim p_{\boldsymbol{\theta^-}}}{\mathbb{E}} K_{\phi}(\hat{\x},\hat{\x}) \bigg)-\\
        &\frac{d}{2m\epsilon^2}\Bigg[\underset{\hat{\x} \sim p_{\boldsymbol{\theta^+}}}{\mathbb{E}} K_{\phi}(\hat{\x},\X) \Big(\I+\frac{d}{m\epsilon^2} K_{\phi}(\X,\X)\Big)^{-1} K_{\phi}(\X,\hat{\x}) \\
        &- \underset{\hat{\x} \sim p_{\boldsymbol{\theta^-}}}{\mathbb{E}} K_{\phi}(\hat{\x}, \X) \Big(\I+\frac{d}{m\epsilon^2} K_{\phi}(\X,\X) \Big)^{-1} K_{\phi}(\X,\hat{\x}) \Bigg],
    \end{split}
\end{equation}
where $\X \in \mathbb{R}^{d\times m}$ is a matrix of generated sample column vectors from $p_{\boldsymbol{\theta}}$ and $\T=[\X, \Y]$ is the concatenated total sample matrix, where $\Y \in \mathbb{R}^{d\times m}$ is a matrix of real sample column vectors from $p^*$. The gradient is exact when $m\rightarrow \infty$, and in practice we estimate this gradient via sampling from $p^*$, $p_{\boldsymbol{\theta}}$, $p_{\boldsymbol{\theta^+}}$ and $p_{\boldsymbol{\theta^-}}$. In the case of a RBF kernel, the term $\underset{\x\sim p_{\boldsymbol{\theta^+}}}{\mathbb{E}}K_{\phi}(\hat{\x}, \hat{\x})-\underset{\x \sim p_{\boldsymbol{\theta^-}}}{\mathbb{E}}K_{\phi}(\hat{\x}, \hat{\x})=1-1=0$. 

All expectation terms in the loss function gradients above can be approximated via sampling during training, unlike the KL divergence or maximum likelihood loss where explicit knowledge of the intractable $p_{\theta}(\x)$ of the quantum generator is required (hence they are not efficient loss functions).
\section{Numerical Simulations} \label{sec:numerical_sims}
We simulate the quantum circuit classically using a custom built quantum simulator based on Numpy and the Scipy sparse matrices package (scipy.sparse). For the adversarial DNN, we build a simple feed-forward network with PyTorch. The Adam optimizer is used for optimizing both the quantum generator and the classical adversarial DNN parameters. All experiments are run on the Harvard FASRC Cluster \cite{FASRC}.

\begin{center}
\begin{figure}
    \centering
    \includegraphics[scale=0.17]{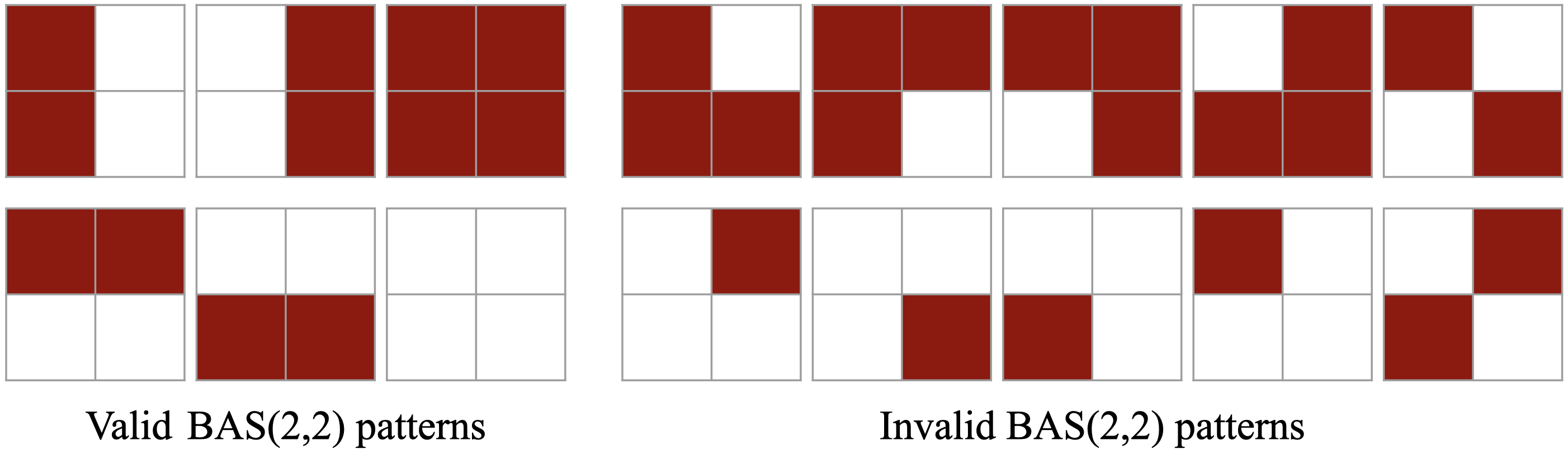}
    \caption{BAS$(2,2)$ patterns modeled by $4$ qubits. The pixel is on (resp. off) if the qubit is measured to be $1$ (resp. $0$)}
    \label{fig: BAS}
\end{figure}
\end{center}

\paragraph{Dataset} Simple patterns such as vertical and horizontal lines are fundamental elements of visual recognition \cite{cortex}. We use the Bars and Stripes (BAS) dataset (Fig. \ref{fig: BAS}), which is a binary-pixel dataset consists of bars (vertical) and stripes (horizontal), a standard for testing quantum generative models due to the limited scaling of classical simulations. For $h \times w$ pixels, the number of valid BAS patterns is $N_\text{BAS}= 2^h +2^w-2$. The target/data distribution, $p(x)=\frac{1}{N_\text{BAS}}$, iff $x$ is a valid BAS image. We choose two BAS datasets, with $h \times w=2\times 2$ and $2\times 3$ respectively. The $2\times 2$ (resp. $2\times 3$) BAS distribution has a total probability space of $2^4=16$ (resp. $2^6=64$) bitstrings, in which there are $6$ (resp. $10$) equal height modes, each corresponding to a valid binary image pattern (fig). The number of qubits needed to model the BAS distribution is equal to the total number of pixels.
\paragraph{Architectures} For the quantum circuit generator (G), we build a $4$ or $6$-layer parametrized quantum circuit respectively for the $2\times 2$ and $2\times 3$ BAS datasets, where the first layer is a single $R_y$ rotation layer with $n=w \times h$ parameters, and each of other layers $l\neq 1$ is composed of $3$ single-qubit rotations for each qubit $j$: $\{\theta_l^{j, 1}, \theta_l^{j, 2}, \theta_l^{j, 3}\}$ and pair-wise nonparametric entanglement gates (CNOT). There are $40$ and $60$ total parameters for each respective dataset. To prevent overpowering the generator, the adversarial DNN (D) simply contains 1 hidden layer, and an additional scoring layer is added for the non-saturating GAN loss term (the penultimate layer outputs are used for the MCR$^2$ second moment matching):
\begin{itemize}
    \item \textbf{Interpolated non-saturating GAN loss and MCR$^2$:} 
    
    Linear($\textit{input size}, 4$)-BatchNorm-ReLU\\
    Linear($4, 4$)-BatchNorm-ReLU\\
    Linear($4,2$)-BatchNorm\\
    Linear($2,1$)-Sigmoid\\
    \item \textbf{Deep Kernel MMD loss:} 
    
    Linear($\textit{input size}, 4$)-BatchNorm-ReLU\\
    Linear($4,4$)-BatchNorm-ReLU\\
    Linear($4,2$)-BatchNorm.
\end{itemize}

\paragraph{Performance Metric} All models are trained till convergence, and we present the results of the converged model parameters with the best learning rates for G and D after a grid search. We evaluate the performance of the model by the Total Variation (TV) metric:
\begin{equation}
    T V\left(p_{\theta}, p^* \right)=\frac{1}{2} \sum_{\boldsymbol{x}}\left|p_{\theta}(\boldsymbol{x})-p^* (\boldsymbol{x})\right|,
\end{equation}
where $p_\theta$ is the probability density of the trained model, and $p^*$ is the data probability density (target distribution). For the quantum Generator circuit, we have direct access to the $p_\theta(x)$ values from the quantum simulator, and we only use this exact $p_\theta(x)$ for evaluating TV. During training, we strictly sample limited batch sizes from $p_\theta(x)$, estimate the gradient, and update the quantum circuit parameters. The adversarial neural network also sees limited batch sizes during training, unless otherwise noted.
\\

\begin{figure*}
\centering
  \includegraphics[scale=0.150]{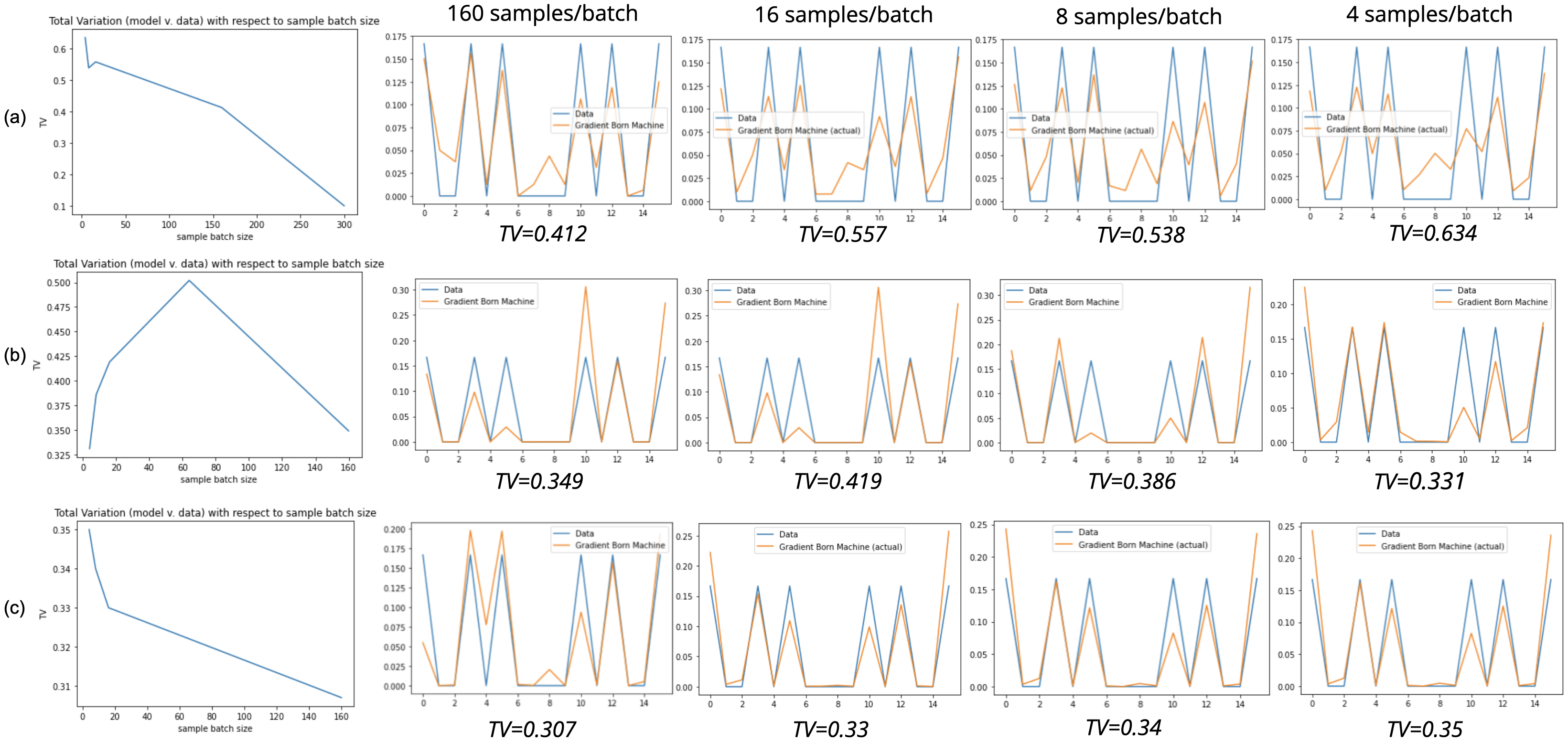}
    \caption{The blue peaks correspond to valid patters and the orange peaks are the exact learnt model distributions. (a) MMD-RBF: large noises are observed with small bath sizes. (b) GAN-NS: invalid image patterns are avoided but mode collapse is apparent. (c) GAN-MCR$^2$: some mode collapse is alleviated. }
    \label{fig:batch}
\end{figure*}

\begin{figure*}
\centering
  \includegraphics[scale=0.15]{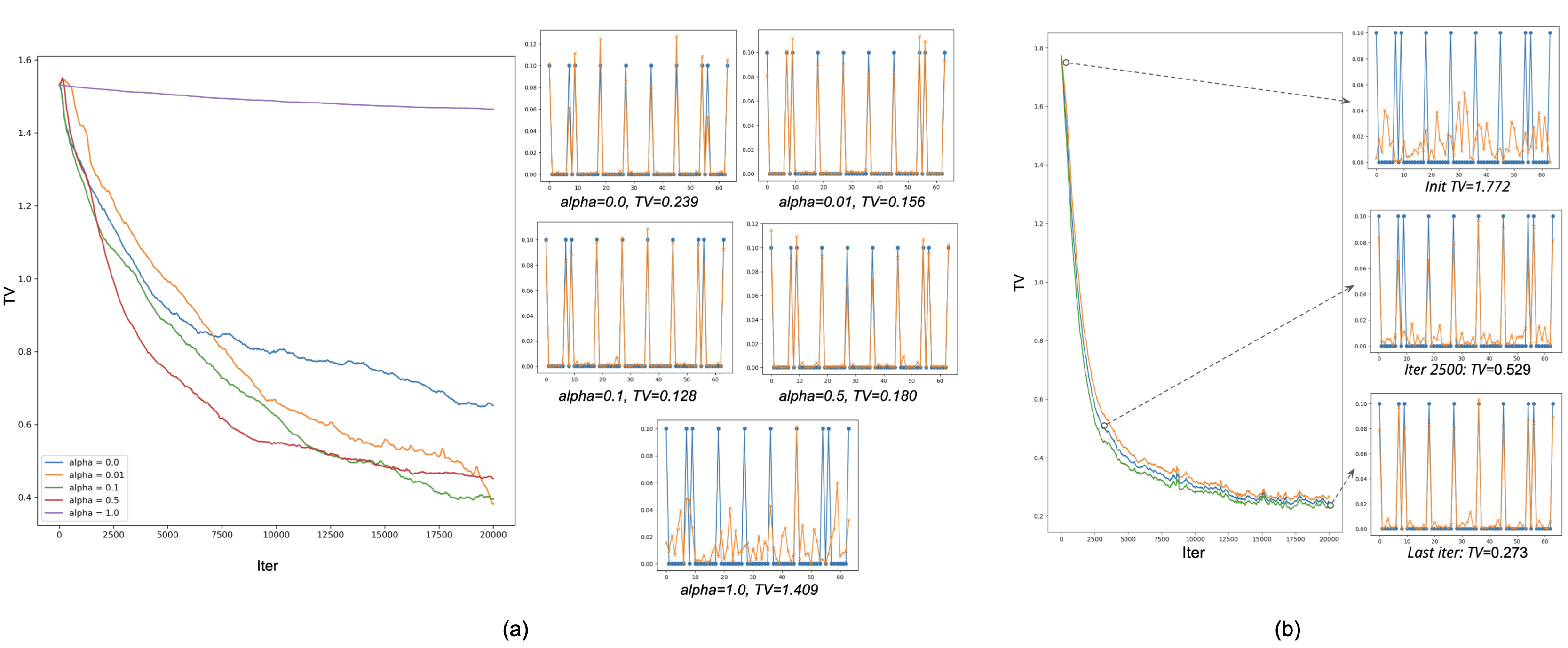}
    \caption{(a) The average TV of QCBM with the \textbf{Inter-NS-MCR$^2$} scheme at various interpolation factor $\alpha$. The good $\alpha$ values are $0.01$ or $0.1$ We also display the final learnt model distributions whose TVs are the lowest out of the $20$ runs. Mode collapse is apparent when $\alpha=0$, corresponding to the non-saturating GAN scheme. b) The training of the QCBM with the \textbf{DNN-MMD} scheme, the blue line is the average TV, and the orange and green lines are one standard deviation above and below the average of $20$ runs. We also display the learnt model distributions at three particular stages during training, from a randomly selected training instance from the $20$ runs.}
    \label{fig:moment}
\end{figure*}

\begin{figure*}[h!]
\centering
    \includegraphics[scale=0.165]{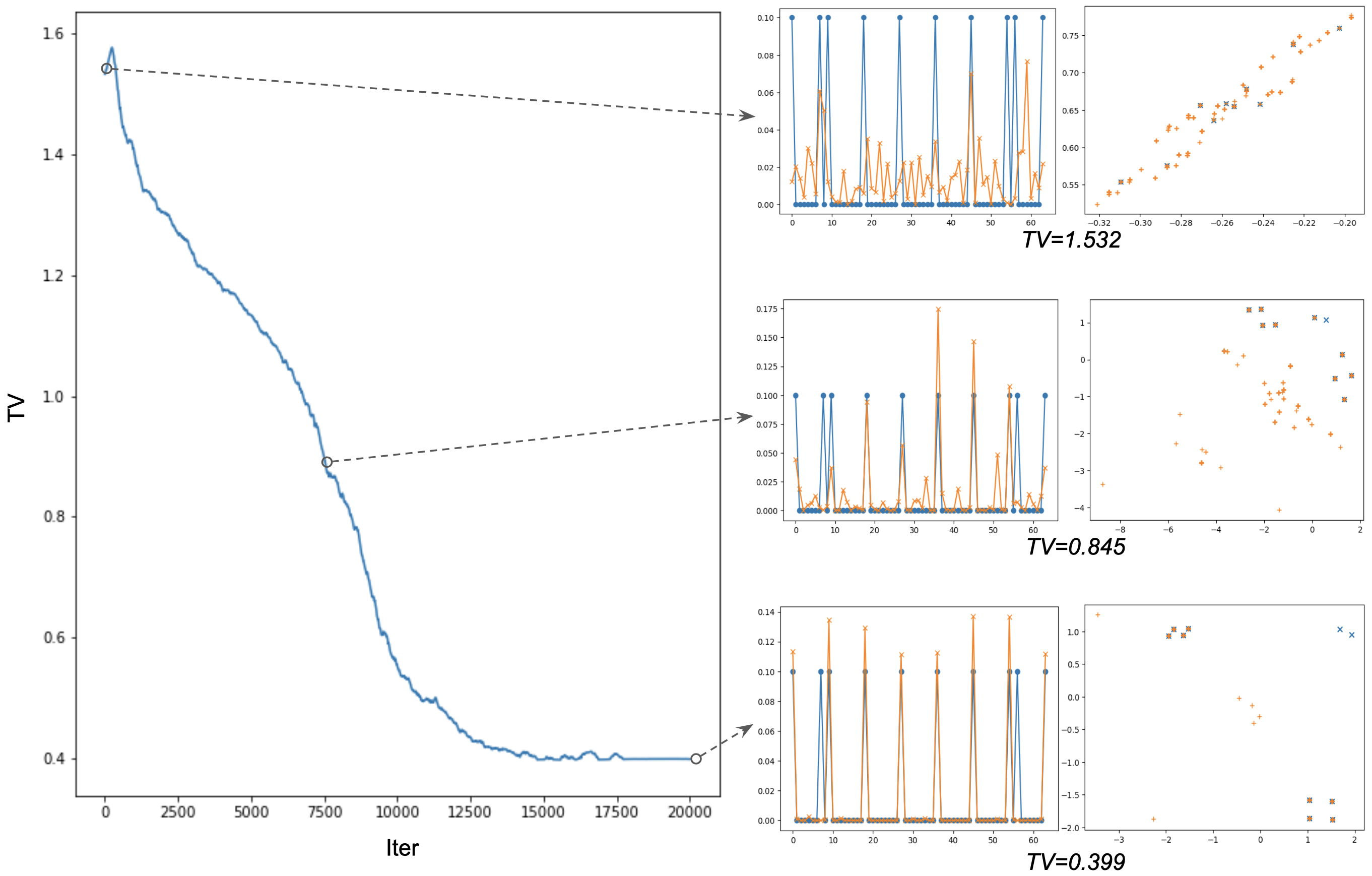}
    \caption{\textbf{QCBM+GAN-NS}: Visualization of the feature space (right column) juxtaposed with the learnt probability distribution (mid column) on the BAS(2,3) dataset, from a particular instance \textbf{using only the non-saturating GAN loss} with an adversarial NN discriminator. Initially, the discriminator cannot discriminate generated samples (orange) from real (blue) samples, so their feature vectors lie in two overlapping Gaussian-like subspaces. During training, the discriminator gradually tells apart generated samples from real samples, mapping them to feature vectors in two clusters. In the end, mode collapse occurs as the quantum generator focuses on a few modes, completely missing two modes (bottom right figure). We later show that this can be alleviated by actively matching the generated sample feature vectors to real sample feature vectors up to the second moment, forcing the quantum generator to output samples with more diverse feature representations.}
    \label{fig:NS-GAN-features}
\end{figure*}

\begin{figure*}[h!]
\centering
    \includegraphics[scale=0.159]{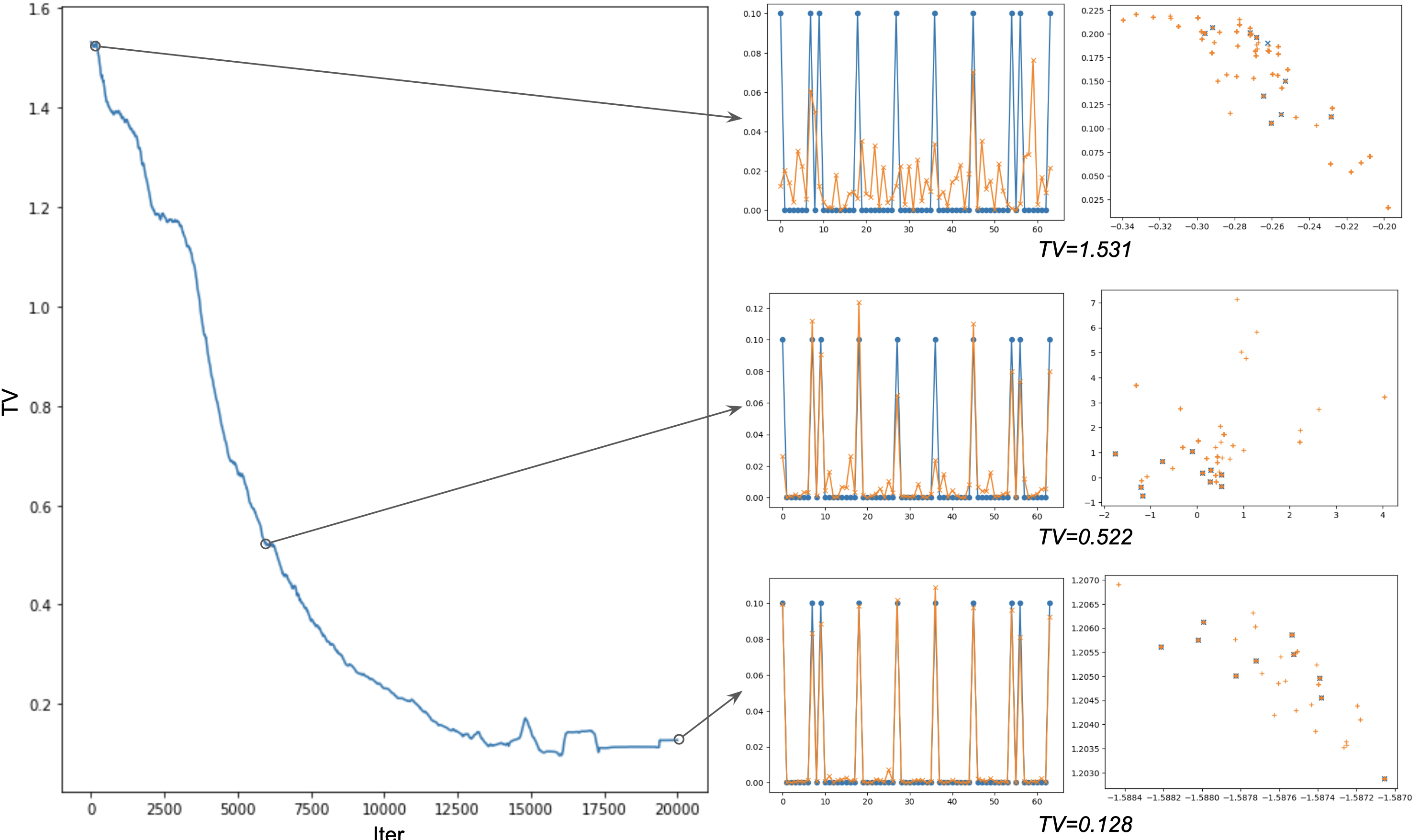}
    \caption{\textbf{QCBM+Inter-NS-MCR$^2$}: Visualization of the feature space (right column) juxtaposed with the learnt probability distribution (mid column) on the BAS(2,3) dataset, from a particular instance \textbf{using the interpolated MCR$^2$ loss with non-saturating GAN loss} with an adversarial DNN feature mapper. Initially, the discriminator cannot discriminate generated samples (orange) from real (blue) samples, so their feature vectors lie in two overlapping Gaussian-like subspaces. Mode collapse is alleviated because the MCR$^2$ forces the quantum generator to output samples with more diverse feature representations, which lie in (approximately) the same Gaussian-like subspace as the real sample vectors (bottom right figure).}
    \label{fig:inter-NS-MCR-features}
\end{figure*}

\subsection{Batch Size Studies}
With the same quantum circuit generator (G), there are three QCBM models under consideration, with or without an adversarial DNN (D):
\begin{enumerate}
    \item \textbf{MMD-RBF} No adversarial DNN, Gaussian RBF MMD loss directed applied to the data space \cite{dif_born}.
    \item \textbf{GAN-NS} Adversarial DNN discriminator scorer, non-sasurating GAN loss.
    \item \textbf{GAN-MCR$^2$} Adversarial DNN feature mapper, second moment matching through the MCR$^2$ loss.
\end{enumerate}
As a control (for this section only), we give the QCBM full knowledge of the data distribution $p^*$, instead of approximating $p^*$ via sampling from data. This is done by keeping track of the full probability densities $p^*(\y)$ and rewriting all $\underset{\y\sim p^*}{\mathbb{E}}f(\y)$ terms as $\sum_{\y} p^*(\y) f(\y)$, for any given $f(\cdot)$ appearing in the loss functions and loss gradients. For all visualizations, the blue peaks correspond to the probability densities of valid BAS patterns (we connect the discrete values for better visualization), and for invalid BAS patterns the densities are $0$. After training, the learnt distribution $p_{\theta}(\x)$ is directly accessed from the classical quantum circuit simulator, and the discrete values are plotted and connected by an orange line. We train the same quantum circuit architecture starting from the same randomly initialized circuit parameters (producing a TV value of $1.064$). Each model is trained until convergence with the best found learning rates (1e-3 for D and G in all cases, except for GAN-NS with batch size of 4, which uses 1e-4 for D). For both GAN-NS and GAN-MCR$^2$, D is updated twice for each G update. The TV metric value with respect to the number of samples per batch is shown in Fig. \ref{fig:batch} (a), left column.

Under small numbers of generated samples, the learning performances of all three models are negatively affected. For MMD-RBF, large noises are observed in the learnt distribution (orange line), and the TV value worsens almost linearly with decreasing batch sizes. For GAN-NS, no correlation is observed beteen model performance and batch size. Noises are surpressed in GAN-NS, and the quantum G is able to avoid generating invalid BAS patterns but with serious mode collapses, confirming with the observation in \cite{adv_Born}, where the authors trained on the $3 \times 3$ BAS dataset with a large batch size of $512$. For GAN-MCR$^2$, the performance decreases with smaller batch sizes, and mode collapse is alleviated to some extend but not consistently. MCR$^2$ is successful in achieving orthogonal subspace-like feature representations ($2$-dimensional feature space visualization in Appendix \ref{Appendix-figures} Fig. \ref{fig: subspace}). Although MCR$^2$ produces structured feature representations, \textbf{matching up to the second moment alone does not completely solve mode collapse}.
Observing a trade-off between noise and mode collapse in the learnt distribution, we tried a two-step train-then-fine-tune scheme: training the GAN-NS until convergence, which brings the quantum model parameters to an advantageous point where invalid patterns are avoided, then followed by fine-tuning the circuit with MMD-RBF, using infinite moment matching to balance out the generated modes. We select the worst-learnt circuit in the above GAN-NS experiments (with a batch size=$64$ at TV=$0.502$) and fine-tune it with the Gaussian kernel MMD loss, with a learning rate of 1e-4 and a batch size of $4$. After another $2000$ iterations, the model produces a good TV=$0.089$ (Appendix \ref{Appendix-figures} Fig. \ref{fig: finetune}). This ad hoc scheme supports our hypothesis that mixing the class probability estimation loss and moment matching loss (which we explore next) could be a viable mediation to mode collapse in QCBMs.

\subsection{Moment Matching Studies}
In this section we keep the batch size at 4 and benchmark different moment-matching schemes on the $2\times 3$ BAS dataset. Both G and D only use $4$ samples to approximate their gradients at each update. D is updated twice per G update. Two schemes are considered:
\begin{enumerate}
    \item \textbf{Inter-NS-MCR$^2$} Adversarial DNN, interpolated generator loss between MCR$^2$ second moment matching and non-saturating GAN loss (\ref{interpolate})
    \item \textbf{DNN-MMD} Adversarial DNN, infinite moment matching through MMD loss in low-dimensional feature space.
\end{enumerate}

In the case of Inter-NS-MCR$^2$, for each value of the interpolation factor $\alpha\in\{0.0, 0.01, 0.1, 0.5, 1.0\}$, we run the experiment $20$ times with a grid search for the best G and D learning rates. We present the average TV measures versus iteration number (of G) with the best found learning rates. The best $\alpha$ values are $0.01$ and $0.01$ (Fig. \ref{fig:moment} (a)). When $\alpha=0$ (corresponding to the non-saturating GAN loss), mode collapse happens, and when $\alpha$ becomes too large, the generator fails to learn. This interpolation between MCR$^2$ and the GAN loss reveals the intricate interplay between two statistical assumptions: probability estimation and moment matching, and it would be interesting to study where exactly the best balance sits. For DNN-MMD, replacing the non-saturating GAN loss and the MCR$^2$ loss all together, we run the experiment $20$ time for each combination of learning rates. The results are presented in Fig. \ref{fig:moment} (b). Both Inter-NS-MCR$^2$ and DNN-MMD resist mode collapse. However DNN-MMD does not yield good TV scores (sometimes even worse than the NS-GAN case without second moment matching) due to the presence of larger noise (incorrectly assigning probabilities to invalid BAS patterns), revealing a trade-off between the balance of relative modes and the total variation of densities. In the practical sense, Inter-NS-MCR$^2$ requires searching for a good $\alpha$ interpolation value, making DNN-MMD a more versatile scheme. For this particular BAS dataset, moment matching is indeed effective against mode collapse and can produce stable, convergent training.

Delving deeper into the feature representations that lead to mode collapse, we visualize the $2$-dimensional feature representations of $500$ samples from the generated and the real distributions for a particular instance of training a QCBM with the \textbf{GAN-NS} scheme (where only the class probability estimation loss is considered, without any moment matching) (Fig. \ref{fig:NS-GAN-features}). At convergence, the quantum generator focuses on a few modes, completely missing two peaks, which is attributed to the fact that the generated sample features do not completely match the statistics of those of the real samples (Fig. \ref{fig:NS-GAN-features} bottom right sub-figure). As a comparison, we show that by penalizing this mismatch of feature statistics up to the second order (the \textbf{Inter-NS-MCR$^2$} scheme with an $\alpha=0.1$ for MCR$^2$), at convergence the generated sample feature vectors align with the real sample feature vectors in a Gaussian-like subspace (Fig. \ref{fig:inter-NS-MCR-features} bottom right sub-figure). Hence \textbf{mode collapse is alleviated as a joint result of class probability estimation loss and the MCR$^2$ loss}.

\section{Conclusion} \label{sec:conclusion}
We studied the performance of various QCBM-based quantum generative models under limited sample sizes and experimentally confirmed that a quantum-neural network hybrid formulation allows small batch training. To address the mode collapse issue in QCBMs, we studied various ways of incorporating moment matching into training QCBMs. We introduced the Maximal Coding Rate Reduction (MCR$^2$) into the field of quantum machine learning for the first time. Namely, we used MCR$^2$ as an add-on regularizer to match the generated data's feature representations to those of the real data up to the second moment. When combined with the non-saturating GAN loss (based on class probability estimation), MCR$^2$ is able to help resist mode collapse and stabilize training. In addition, we tested the performance of the MMD-GAN scheme in the context of QCBM and showed that it alleviates mode collapse but with a cost of higher noise in the learnt distribution. Future work include deeper investigations into the interaction between quantum circuits and the neural network in the adversarial setting as well as better optimization schemes other than stochastic gradient descent.

{\small
\bibliographystyle{ieee_fullname}
\bibliography{egbib}
}

\newpage
\onecolumn

\appendix

\section{MCR$^2$ Quantum Circuit Gradient Derivations (Explicit Features)} \label{Appendix-explicit}
In the case when we have explicit mapping  from data space to feature space, i.e., $Z = \phi(X)\in \mathbb{R}^{n\times m}$, where $\phi(\cdot)$ can be a neural network parameterized by $\theta$. To avoid confusion, in this section we denote the feature matrix $Z=f_{\theta}(X)\in \mathbb{R}^{n\times m}$. Not requiring the inner product form (as in the later section), the quantum gradient is computationally less expensive. The MCR$^2$ metric is:
\begin{equation}
    \begin{split}
        \Delta &\lim_{m\rightarrow \infty} R([\phi(\X),\phi(\Y)])=\\
        &\underbrace{\frac{1}{2}\log\det(\I + \frac{d}{2\epsilon^2}\sum_{\x} p_{\boldsymbol{\theta}}(\x)\phi(\x)\phi(\x)^T+\frac{d}{2\epsilon^2}\sum_{\y} p_{\Y}(\y)\phi(\y)\phi(\y)^T)}_{\mathcal{L}_A}\\
        &-\underbrace{\frac{1}{4}\log\det(\I + \frac{d}{\epsilon^2}\sum_{\x} p_{\boldsymbol{\theta}}(\x)\phi(\x)\phi(\x)^T)}_{\mathcal{L}_B}-\frac{1}{4}\log\det(\I + \frac{d}{\epsilon^2}\sum_{\y} p_{\Y}(\y)\phi(\y)\phi(\y)^T)\\
        &= \mathcal{L}_{\text{MCR}^2},
    \end{split}
\end{equation}
and directly using the fact that $\frac{\partial \log\det(A)}{\partial A}=A^{-1}$, we have:
\begin{equation}
    \begin{split}
        \frac{\partial \mathcal{L}_A}{\partial p_\theta (\hat{\x})}=\frac{1}{2}<(\I + \frac{d}{2\epsilon^2}\sum_{\x} p_{\boldsymbol{\theta}}(\x)\phi(\x)\phi(\x)^T+\frac{d}{2\epsilon^2}\sum_{\y} p_{\Y}(\y)\phi(\y)\phi(\y)^T)^{-1}, \phi(\hat{\x})\phi(\hat{\x})^\top>,
    \end{split}
\end{equation}
and similarly:
\begin{equation}
    \begin{split}
        &\frac{\partial \mathcal{L}_B}{\partial p_\theta (\hat{\x})}=\frac{1}{4}<(\I + \frac{d}{\epsilon^2}\sum_{\x} p_{\boldsymbol{\theta}}(\x)\phi(\x)\phi(\x)^T)^{-1}, \phi(\hat{\x})\phi(\hat{\x})^\top>,
    \end{split}
\end{equation}
where $<A,B>$ denotes the element-wise dot product between matrices $A$ and $B$.

Therefore,
\begin{equation}
    \begin{split}
        \frac{\partial \mathcal{L}_A}{\partial \hat{\theta}}&=\sum_{\x} \frac{\partial \mathcal{L}_A}{\partial p_{\boldsymbol{\theta}}(\x)}\frac{\partial p_{\boldsymbol{\theta}}(\x)}{\partial \hat{\theta}}=\sum_{\x} \frac{\partial \mathcal{L}_A}{\partial p_{\boldsymbol{\theta}}(\x)}\frac{p_{\theta^+}(\x)-p_{\theta^-}(\x)}{2}\\
        &=\frac{1}{4}\mathbb{E}_{\hat{\x}\sim p_{\theta^+}}<(\I + \frac{d}{2\epsilon^2}\sum_{\x} p_{\boldsymbol{\theta}}(\x)\phi(\x)\phi(\x)^T+\frac{d}{2\epsilon^2}\sum_{\y} p_{\Y}(\y)\phi(\y)\phi(\y)^T)^{-1}, \phi(\hat{\x})\phi(\hat{\x})^\top>\\
        &-\frac{1}{4}\mathbb{E}_{\hat{\x}\sim p_{\theta^-}}<(\I + \frac{d}{2\epsilon^2}\sum_{\x} p_{\boldsymbol{\theta}}(\x)\phi(\x)\phi(\x)^T+\frac{d}{2\epsilon^2}\sum_{\y} p_{\Y}(\y)\phi(\y)\phi(\y)^T)^{-1}, \phi(\hat{\x})\phi(\hat{\x})^\top>,
    \end{split}
\end{equation}
and
\begin{equation}
    \begin{split}
        \frac{\partial \mathcal{L}_B}{\partial \hat{\theta}}&=\sum_{\x} \frac{\partial \mathcal{L}_B}{\partial p_{\boldsymbol{\theta}}(\x)}\frac{\partial p_{\boldsymbol{\theta}}(\x)}{\partial \hat{\theta}}=\sum_{\x} \frac{\partial \mathcal{L}_B}{\partial p_{\boldsymbol{\theta}}(\x)}\frac{p_{\theta^+}(\x)-p_{\theta^-}(\x)}{2}\\
        &=\frac{1}{8}\mathbb{E}_{\hat{\x}\sim p_{\theta^+}}<(\I + \frac{d}{\epsilon^2}\sum_{\x} p_{\boldsymbol{\theta}}(\x)\phi(\x)\phi(\x)^T)^{-1}, \phi(\hat{\x})\phi(\hat{\x})^\top>\\
        &-\frac{1}{8}\mathbb{E}_{\hat{\x}\sim p_{\theta^-}}<(\I + \frac{d}{\epsilon^2}\sum_{\x} p_{\boldsymbol{\theta}}(\x)\phi(\x)\phi(\x)^T)^{-1}, \phi(\hat{\x})\phi(\hat{\x})^\top>,
    \end{split}
\end{equation}
where $\langle A, B \rangle$ denotes the dot product of elements of matrices $A$ and $B$.

\section{MCR$^2$ Quantum Circuit Gradient Derivations (Implicit Features via Kernel)} \label{Appendix-kernel}
In the simplest case (compact distributions $ p_X$ and $ p_Y$), given two groups of samples, $\X=\{x\}\sim p_X$ and $\Y=\{y\}\sim p_Y$, and if we treat each sample as a column vector and construct $\X$ and $\Y$, both are $\mathcal{R}^{d\times m}$, where $d$ is the dimension of features and $m$ is the batch size, we define the MCR (second order) ``distance'' between $\X$ and $\Y$ as:
\begin{equation}
\label{sample}
    \begin{split}
        \Delta R([\X,\Y])=&\frac{1}{2}\log\det(\I + \frac{d}{2m\epsilon^2}\X\X^T+\frac{d}{2m\epsilon^2}\Y\Y^T)\\
        &-\frac{1}{4}\log\det(\I + \frac{d}{m\epsilon^2}\X\X^T)-\frac{1}{4}\log\det(\I + \frac{d}{m\epsilon^2}\Y\Y^T).
    \end{split}
\end{equation}
This metric is $0$ iff $\frac{1}{m}\X\X^T=\frac{1}{m}\Y\Y^T$, i.e., the sample covariance matrices (given the same batch sizes) are equal. 

For the case when the sample size $m\rightarrow \infty$, we have:
\begin{equation}
    \lim_{m\rightarrow \infty} \frac{d}{m\epsilon^2}\X\X^T=\frac{d}{\epsilon^2}\sum_{\x} p_{\X}(\x)\x\x^T.
\end{equation}

Therefore:
\begin{equation}
    \begin{split}
        \Delta \lim_{m\rightarrow \infty} R([\X,\Y])=&\frac{1}{2}\log\det(\I + \frac{d}{2\epsilon^2}\sum_{\x} p_{\X}(\x)\x\x^T+\frac{d}{2\epsilon^2}\sum_{\y} p_{\Y}(\y)\y\y^T)\\
        &-\frac{1}{4}\log\det(\I + \frac{d}{\epsilon^2}\sum_{\x} p_{\X}(\x)\x\x^T)-\frac{1}{4}\log\det(\I + \frac{d}{\epsilon^2}\sum_{\y} p_{\Y}(\y)\y\y^T).
    \end{split}
\end{equation}
In the case when $p_X$ and $p_Y$ are Gaussian distributions, matching the above objective (covariances) along with the first moment (mean) will be sufficient for showing $p_X=p_Y$. In practice, $p_X$ and $p_Y$ are intractable, so we can only approximate (\ref{infinite}) by (\ref{sample}) with large enough batch size $m$.
\\

We now consider a reproducing kernel Hilbert space which has an implicit feature mapping $\phi(\cdot)$. By the law of unconscious statistician, $\mathbb{E}_{\phi(\x)\sim p_{\phi(x)}}\phi(\x)\phi(\x)^{\top}=\mathbb{E}_{\x\sim p_{\X}}\phi(\x)\phi(\x)^{\top}$, so the covariance matrix in the kernel feature space can be written as the expectation over $\x\sim p_{\X}$:
\begin{equation}
    \begin{split}
        \Delta \lim_{m\rightarrow \infty} R([\phi(\X),\phi(\Y)])=&\frac{1}{2}\log\det(\I + \frac{d}{2\epsilon^2}\sum_{\x} p_{\X}(\x)\phi(\x)\phi(\x)^T+\frac{d}{2\epsilon^2}\sum_{\y} p_{\Y}(\y)\phi(\y)\phi(\y)^T)\\
        &-\frac{1}{4}\log\det(\I + \frac{d}{\epsilon^2}\sum_{\x} p_{\X}(\x)\phi(\x)\phi(\x)^T)-\frac{1}{4}\log\det(\I + \frac{d}{\epsilon^2}\sum_{\y} p_{\Y}(\y)\phi(\y)\phi(\y)^T).
    \end{split}
\end{equation}
\\

Designating $p_{\X}$ to be the distribution that we want to parameterize via a quantum circuit, which uses a vector of parameters $\boldsymbol{\theta}$ to model $p_{\X}$:
\begin{equation}
\label{kernel_}
    \begin{split}
        \Delta \lim_{m\rightarrow \infty} R([\phi(\X),\phi(\Y)])=&\underbrace{\frac{1}{2}\log\det(\I + \frac{d}{2\epsilon^2}\sum_{\x} p_{\boldsymbol{\theta}}(\x)\phi(\x)\phi(\x)^T+\frac{d}{2\epsilon^2}\sum_{\y} p_{\Y}(\y)\phi(\y)\phi(\y)^T)}_{\mathcal{L}_A}\\
        &-\underbrace{\frac{1}{4}\log\det(\I + \frac{d}{\epsilon^2}\sum_{\x} p_{\boldsymbol{\theta}}(\x)\phi(\x)\phi(\x)^T)}_{\mathcal{L}_B}-\frac{1}{4}\log\det(\I + \frac{d}{\epsilon^2}\sum_{\y} p_{\Y}(\y)\phi(\y)\phi(\y)^T)\\
        &= \mathcal{L}_{\text{MCR}^2}
    \end{split}
\end{equation}
\\

Now we derive the quantum gradient of the above metric with respect to a specific single quantum circuit parameter $\hat{\theta}\in \boldsymbol{\theta}$. First note, by the structure of the above loss function:
\begin{equation}
    \frac{\partial \mathcal{L}}{\partial \hat{\theta}}=\sum_{\x} \frac{\partial \mathcal{L}}{\partial p_{\boldsymbol{\theta}}(\x)}\frac{\partial p_{\boldsymbol{\theta}}(\x)}{\partial \hat{\theta}},
\end{equation}
where 
\begin{equation}
    \frac{\partial p_{\boldsymbol{\theta}}(\x)}{\partial \hat{\theta}}=\frac{1}{2}\left(p_{\theta^{+}}(\x)-p_{\theta^{-}}(\x)\right).
\end{equation}


To be able to express everything by inner products, we use the fact that:
\begin{equation}
    \begin{split}
      \frac{d}{d x} \log \operatorname{det} A(x)= \frac{d}{d x} \operatorname{Tr} \log A(x)= \operatorname{Tr}\left(A(x)^{-1} \frac{d A}{d x}\right).
    \end{split}
\end{equation}
regardless whether matrices $\frac{d A}{d x}$ and $A(x)$ commute or not \footnote{Derivative of matrix logarithm (Stack Exchange Mathematics). url: math.stackexchange.com/questions/2837070/derivative-
of-matrix-logarithm, 2018. Accessed November, 2022}.

Hence,
\begin{equation}
    \begin{split}
      2\frac{\partial \mathcal{L}_A}{\partial p_{\boldsymbol{\theta}}(\hat{\x})}&=\Tr \Big[\Big(\I + \frac{d}{2\epsilon^2}\sum_{\x} p_{\boldsymbol{\theta}}(\x)\phi(\x)\phi(\x)^T+\frac{d}{2\epsilon^2}\sum_{\y} p_{\Y}(\y)\phi(\y)\phi(\y)^T \Big)^{-1} \phi(\hat{\x})\phi(\hat{\x})^T \Big]\\
      &=\lim_{m\rightarrow \infty} \Tr \Big[\Big(\I + \frac{d}{2m\epsilon^2}\phi(\X)\phi(\X)^T+\frac{d}{2m\epsilon^2}\phi(\Y)\phi(\Y)^T \Big)^{-1} \phi(\hat{\x})\phi(\hat{\x})^T \Big]\\
      &=\lim_{m\rightarrow \infty} \Tr \Big[\Big(\I + \frac{d}{2m\epsilon^2}\underbrace{[\phi(\X), \phi(\Y)]}_{M}\underbrace{[\phi(\X), \phi(\Y)]^T}_{M^T} \Big)^{-1} \phi(\hat{\x})\phi(\hat{\x})^T \Big]\\
      &=\lim_{m\rightarrow \infty} \Tr \Big[\Big(\I - \frac{d}{2m\epsilon^2}M\Big(\I+\frac{d}{2m\epsilon^2} M^TM\Big)^{-1} M^T \Big) \phi(\hat{\x})\phi(\hat{\x})^T \Big]\\
      &=\lim_{m\rightarrow \infty} \Tr \Big[\phi(\hat{\x})\phi(\hat{\x})^T - \frac{d}{2m\epsilon^2}M\Big(\I+\frac{d}{2m\epsilon^2} M^TM\Big)^{-1} M^T  \phi(\hat{\x})\phi(\hat{\x})^T \Big]\\
      &=\lim_{m\rightarrow \infty} \Tr \Big[\phi(\hat{\x})\phi(\hat{\x})^T\Big] - \Tr\Big[ \frac{d}{2m\epsilon^2}M\Big(\I+\frac{d}{2m\epsilon^2} M^TM\Big)^{-1} M^T  \phi(\hat{\x})\phi(\hat{\x})^T \Big]\\
      &=\lim_{m\rightarrow \infty} \Tr \Big[\phi(\hat{\x})^T \phi(\hat{\x})\Big] - \Tr\Big[ \frac{d}{2m\epsilon^2}\phi(\hat{\x})^T M\Big(\I+\frac{d}{2m\epsilon^2} M^TM\Big)^{-1} M^T  \phi(\hat{\x}) \Big]\\
      &=\lim_{m\rightarrow \infty} \Big\{\phi(\hat{\x})^T \phi(\hat{\x}) - \frac{d}{2m\epsilon^2}\phi(\hat{\x})^T M\Big(\I+\frac{d}{2m\epsilon^2} M^TM\Big)^{-1} M^T  \phi(\hat{\x}) \Big\}
    \end{split}
\end{equation}
Then, the gradient of $\mathcal{L}_A$ is expressed as:
\begin{equation}
    \begin{split}
        2\frac{\partial \mathcal{L}_A}{\partial \hat{\theta}}&=\sum_{\x} \frac{\partial \mathcal{L}}{\partial p_{\boldsymbol{\theta}}(\x)}\frac{\partial p_{\boldsymbol{\theta}}(\x)}{\partial \hat{\theta}}=\lim_{m\rightarrow \infty}\sum_{\hat{\x}}  \Big[\phi(\hat{\x})^T \phi(\hat{\x})\Big]\frac{\partial p_{\boldsymbol{\theta}}(\x)}{\partial \hat{\theta}}\\
        &-\sum_{\hat{\x}} \Big[ \frac{d}{2m\epsilon^2}\phi(\hat{\x})^T M\Big(\I+\frac{d}{2m\epsilon^2} M^TM\Big)^{-1} M^T  \phi(\hat{\x}) \Big]\frac{\partial p_{\boldsymbol{\theta}}(\x)}{\partial \hat{\theta}}\\
        &=\frac{1}{2}\mathbb{E}_{\x\sim p_{\boldsymbol{\theta^+}}}\phi(\hat{\x})^T \phi(\hat{\x})-\frac{1}{2}\mathbb{E}_{\x \sim p_{\boldsymbol{\theta^-}}}\phi(\hat{\x})^T \phi(\hat{\x})\\
        &-\lim_{m\rightarrow \infty}\frac{1}{2}\Bigg\{\frac{d}{2m\epsilon^2}\mathbb{E}_{\x \sim p_{\boldsymbol{\theta^+}}}\phi(\hat{\x})^T M\Big(\I+\frac{d}{2m\epsilon^2} M^TM\Big)^{-1} M^T  \phi(\hat{\x}) \\
        &- \frac{d}{2m\epsilon^2}\mathbb{E}_{\x \sim p_{\boldsymbol{\theta^-}}}\phi(\hat{\x})^T M\Big(\I+\frac{d}{2m\epsilon^2} M^TM\Big)^{-1} M^T  \phi(\hat{\x}) \Bigg\}.\\
    \end{split}
\end{equation}
Using the similar procedure, we can see the gradient of $\mathcal{L}_B$ is:
\begin{equation}
    \begin{split}
        4\frac{\partial \mathcal{L}_B}{\partial \hat{\theta}}&=\frac{1}{2}\mathbb{E}_{\hat{\x}\sim p_{\boldsymbol{\theta^+}}}\phi(\hat{\x})^T \phi(\hat{\x})-\frac{1}{2}\mathbb{E}_{\hat{\x} \sim p_{\boldsymbol{\theta^-}}}\phi(\hat{\x})^T \phi(\hat{\x})\\
        &-\lim_{m\rightarrow \infty}\frac{1}{2}\Bigg\{\frac{d}{m\epsilon^2}\mathbb{E}_{\hat{\x} \sim p_{\boldsymbol{\theta^+}}}\phi(\hat{\x})^T \phi(\X)\Big(\I+\frac{d}{m\epsilon^2} \phi(\X)^T \phi(\X)\Big)^{-1} \phi(\X)^T  \phi(\hat{\x}) \\
        &- \frac{d}{m\epsilon^2}\mathbb{E}_{\hat{\x} \sim p_{\boldsymbol{\theta^-}}}\phi(\hat{\x})^T \phi(\X)\Big(\I+\frac{d}{m\epsilon^2} \phi(\X)^T \phi(\X)\Big)^{-1} \phi(\X)^T  \phi(\hat{\x}) \Bigg\}.
    \end{split}
\end{equation}
For the case of a RBF kernel $K_{\phi}(\cdot, \cdot)$, the term $\mathbb{E}_{\hat{\x}\sim p_{\boldsymbol{\theta^+}}}\phi(\hat{\x})^T \phi(\hat{\x})-\mathbb{E}_{\hat{\x} \sim p_{\boldsymbol{\theta^-}}}\phi(\hat{\x})^T \phi(\hat{\x})=1-1=0$. The matrix $M^\top M$ is an inner product form and thus can be expressed via a kernel matrix $K_{\phi} \left([\X, \Y], [\X, \Y]\right)$. Similarly, 
\begin{equation}
    \begin{split}
        &M^\top \phi(\hat{\x})=K_{\phi}([\X, \Y], \hat{\x}),\\
        &\phi(\X)^T  \phi(\hat{\x})=K_{\phi}(\X, \hat{\x}),\\
        &\phi(\X)^T \phi(\X)=K_{\phi}(\X, \X).
    \end{split}
\end{equation}

\section{Figures} \label{Appendix-figures}
\begin{figure}[h]
    \includegraphics[scale=0.22]{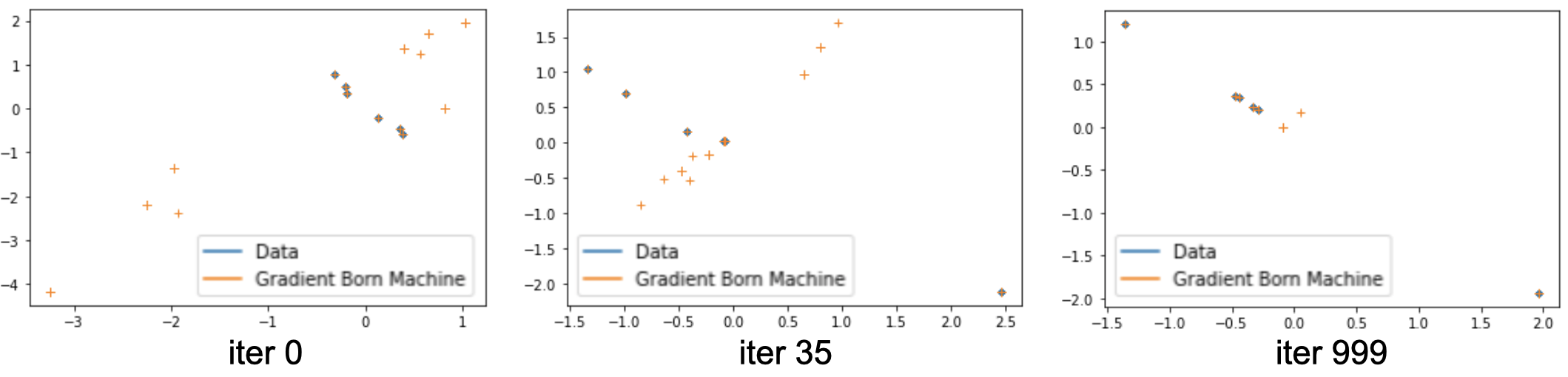}
    \caption{\textbf{QCBM+MCR$^2$}: Visualization of the feature space with on BAS(2,2) dataset, \textbf{using solely MCR$^2$ as the loss function} with an adversarial DNN feature mapper. At early training stages, the Discriminator maps the real and synthesized data points into almost orthogonal linear subspaces (e.g. iter 35). The quantum Generator then learns to align with the real data in later iterations (e.g. iter 999).}
    \label{fig: subspace}
\end{figure}

\begin{figure}[h]
    \includegraphics[scale=0.17]{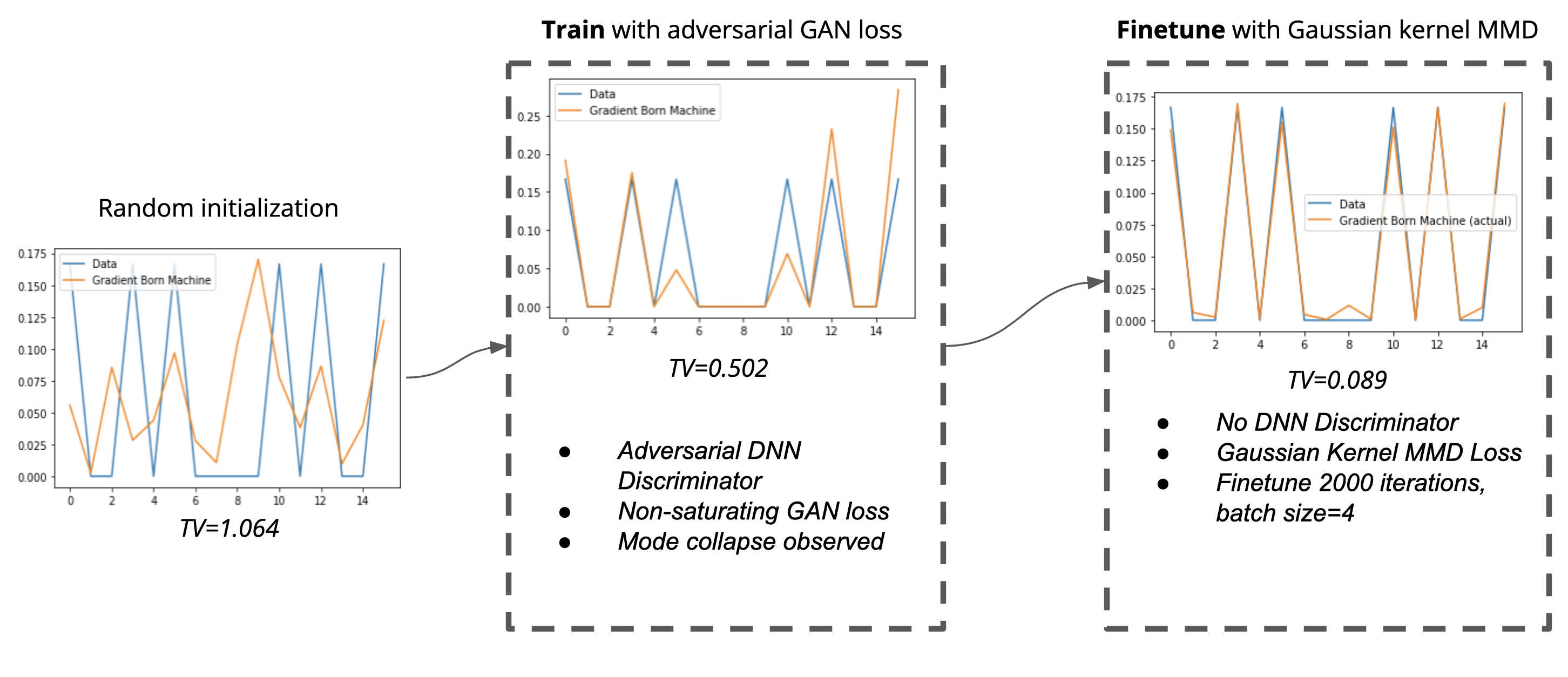}
    \caption{\textbf{Two-step Train then Fine-tune}: With the $2\times2$ BAS dataset, we first train the QCBM with the non-saturating GAN loss with a batch size of $4$ for both G and D, which eliminates the generation of invalid patterns; we then fine-tune the circuit with the Gaussian Kernel MMD loss directly in the image space, with $2000$ iterations under a batch size of $4$ per iteration. The mode collapse is fixed.}
    \label{fig: finetune}
\end{figure}

\begin{figure}[h]
    \includegraphics[scale=0.165]{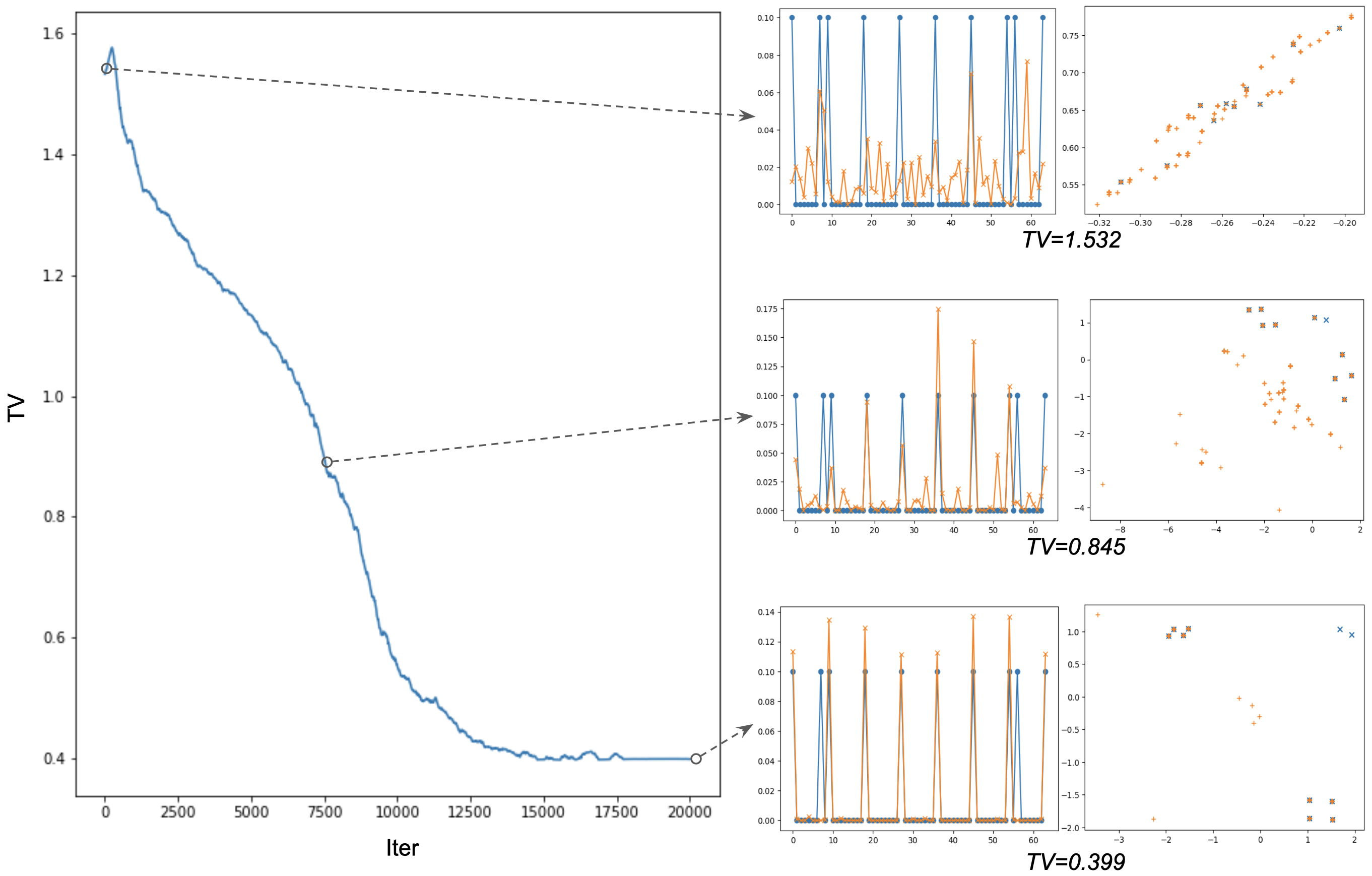}
    \caption{\textbf{QCBM+GAN-NS}: Visualization of the feature space (right column) juxtaposed with the learnt probability distribution (mid column) on the BAS(2,3) dataset, \textbf{using only the non-saturating GAN loss} with an adversarial NN discriminator. Initially, the discriminator cannot discriminate generated samples (orange) from real (blue) samples, so their feature vectors lie in two overlapping Gaussian-like subspaces. During training, the discriminator gradually tells apart generated samples from real samples, mapping them to feature vectors in two clusters. In the end, mode collapse occurs as the quantum generator focuses on a few modes, completely missing two modes (bottom right figure). We later show that this can be alleviated by actively matching the generated sample feature vectors to real sample feature vectors up to the second moment, forcing the quantum generator to output samples with more diverse feature representations.}
\end{figure}

\begin{figure}[h!]
    \includegraphics[scale=0.159]{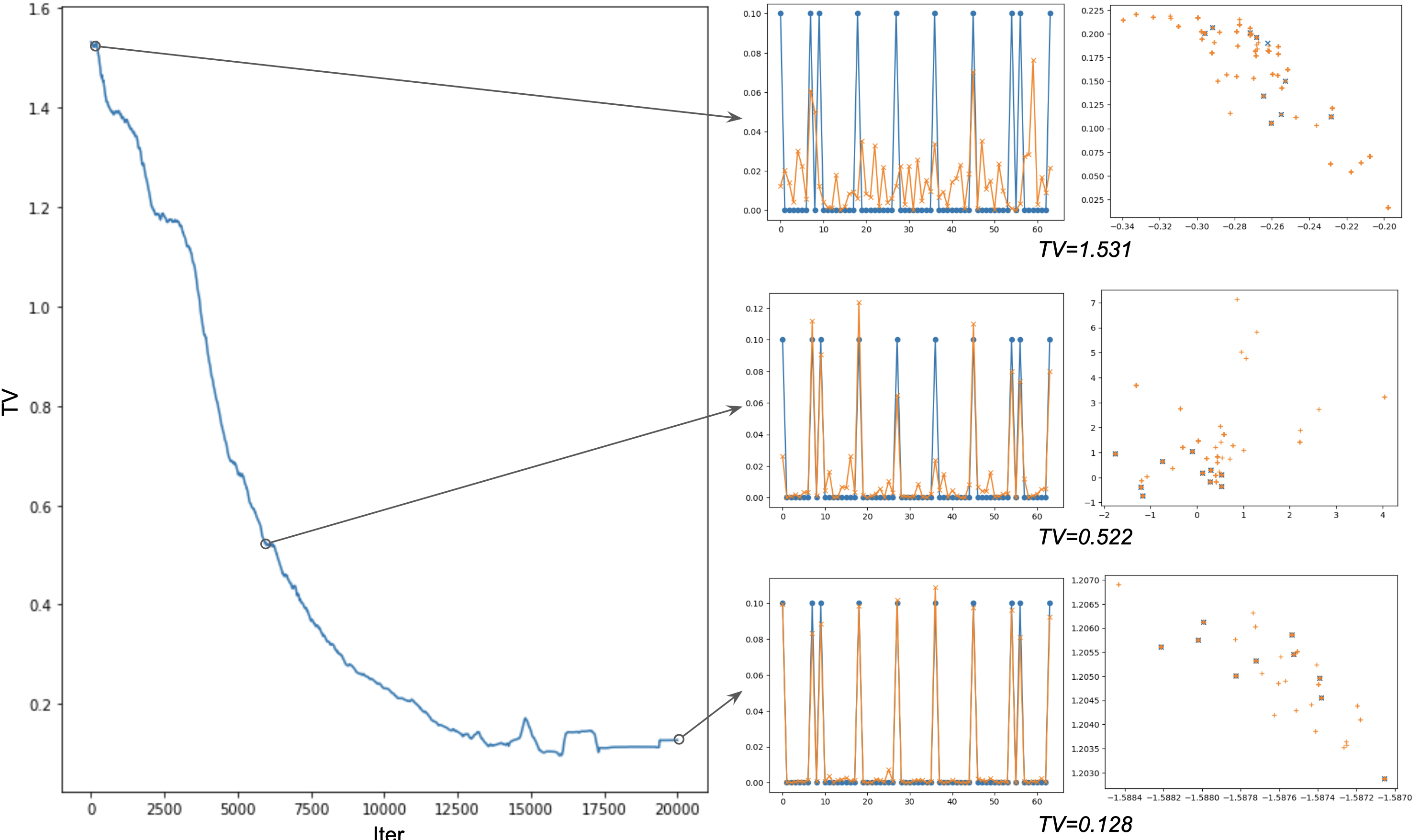}
    \caption{\textbf{QCBM+Inter-NS-MCR$^2$}: Visualization of the feature space (right column) juxtaposed with the learnt probability distribution (mid column) on the BAS(2,3) dataset, \textbf{using the interpolated MCR$^2$ loss with non-saturating GAN loss} with an adversarial DNN feature mapper. Initially, the discriminator cannot discriminate generated samples (orange) from real (blue) samples, so their feature vectors lie in two overlapping Gaussian-like subspaces. Mode collapse is alleviated because the MCR$^2$ forces the quantum generator to output samples with more diverse feature representations, which lie in (approximately) the same Gaussian-like subspace as the real sample vectors (bottom right figure).}
\end{figure}


\begin{figure*}[h!]
\centering
    \includegraphics[scale=0.165]{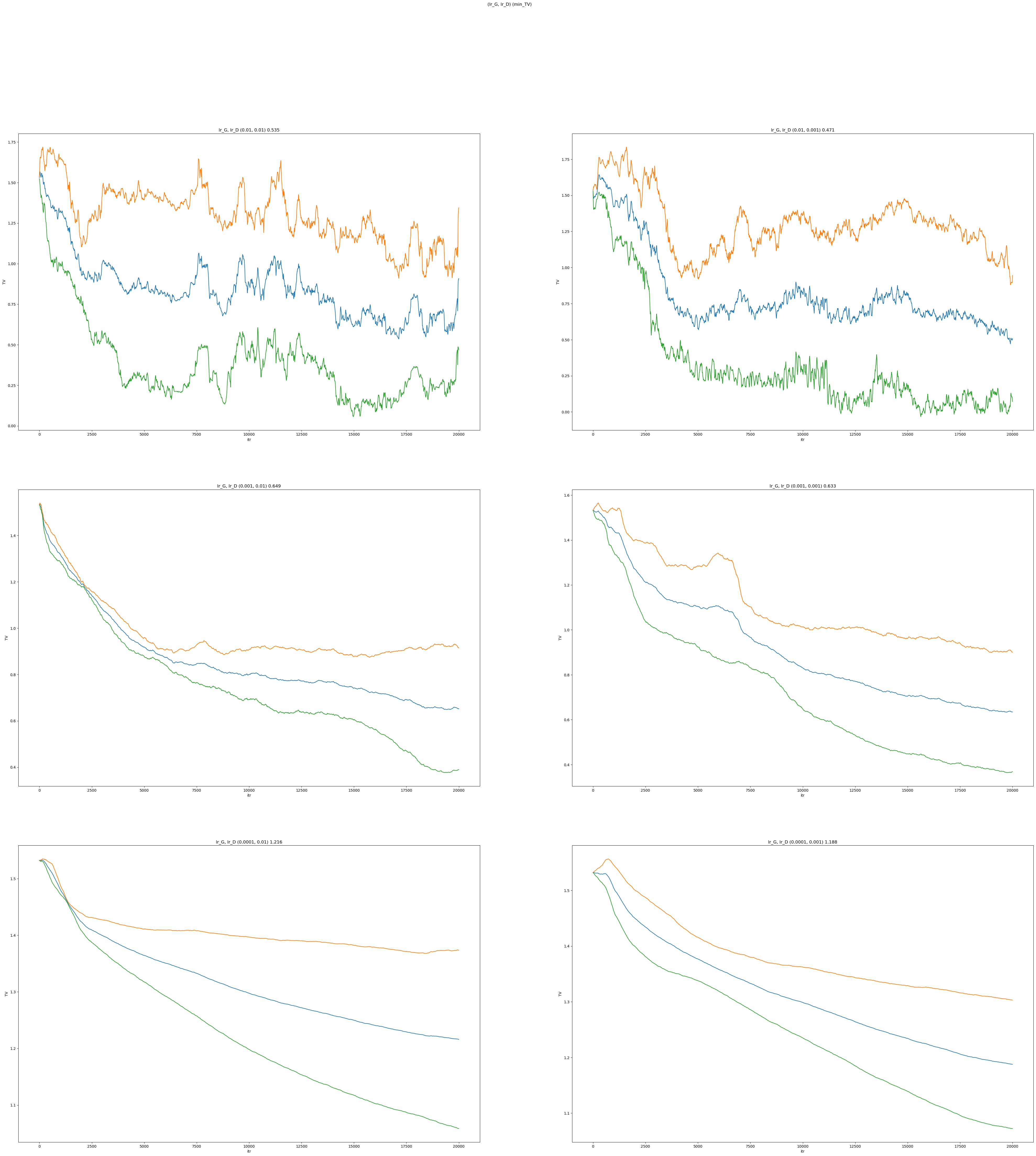}
    \caption{\textbf{QCBM+Inter-NS-MCR$^2$ with $\alpha=0.0$}: The total variation (TV) versus iteration number of the generated distribution learnt by the quantum circuit at various learning rates (lr) for the discriminator (D) and the generator (G). We vary G's lr at different rows ($0.01, 0.001, 0.0001$ top to bottom) and vary D's lr at different columns ($0.01, 0.001$ left to right). The blue line is the average TV value across $20$ runs, and the orange and green lines are one standard deviation above and below the average value.}
\end{figure*}

\begin{figure*}[h!]
\centering
    \includegraphics[scale=0.165]{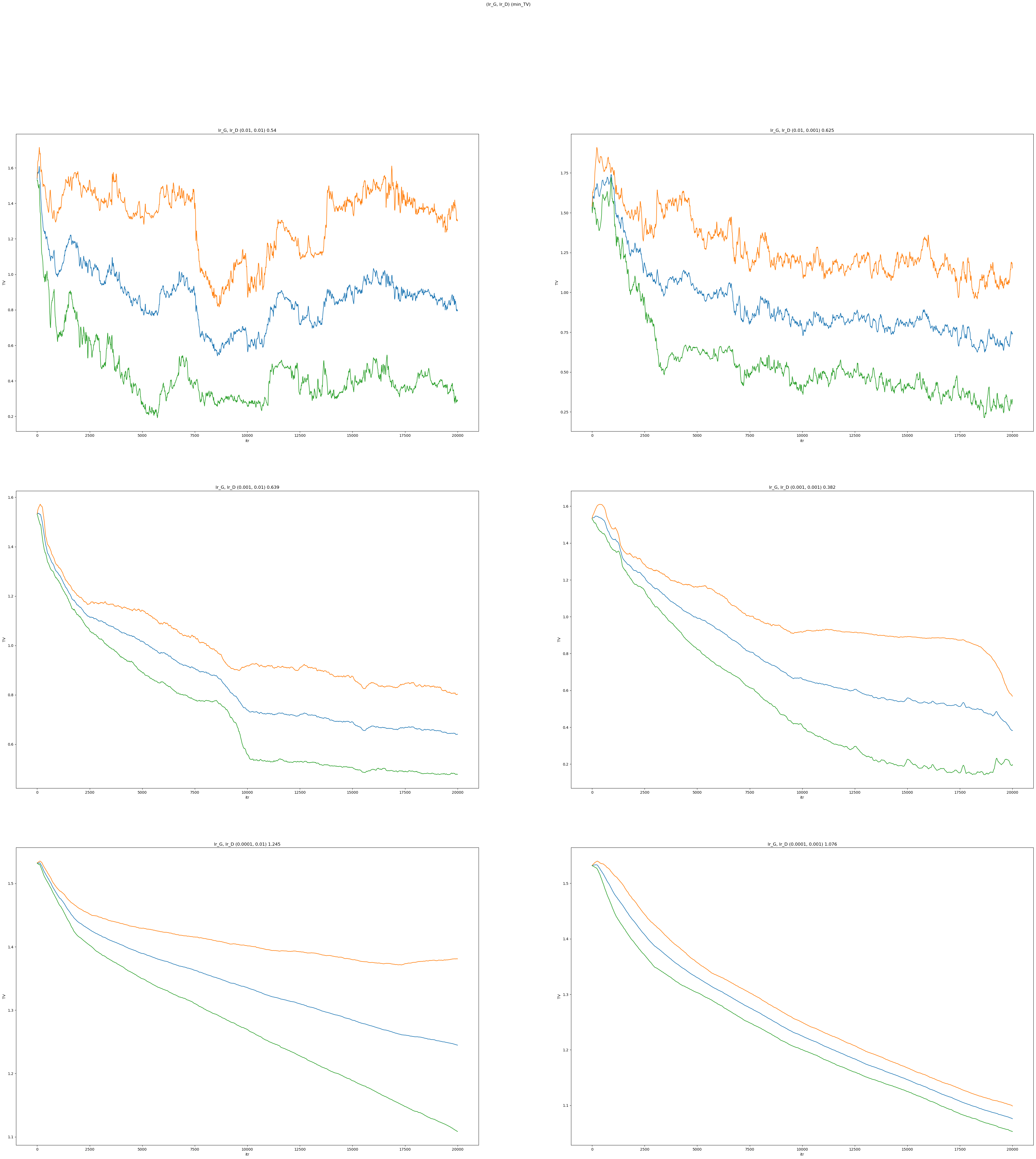}
    \caption{\textbf{QCBM+Inter-NS-MCR$^2$ with $\alpha=0.01$}: The total variation (TV) versus iteration number of the generated distribution learnt by the quantum circuit at various learning rates (lr) for the discriminator (D) and the generator (G). We vary G's lr at different rows ($0.01, 0.001, 0.0001$ top to bottom) and vary D's lr at different columns ($0.01, 0.001$ left to right). The blue line is the average value across $20$ runs, and the orange and green lines are one standard deviation above and below the average value.}
\end{figure*}

\begin{figure*}[h!]
\centering
    \includegraphics[scale=0.165]{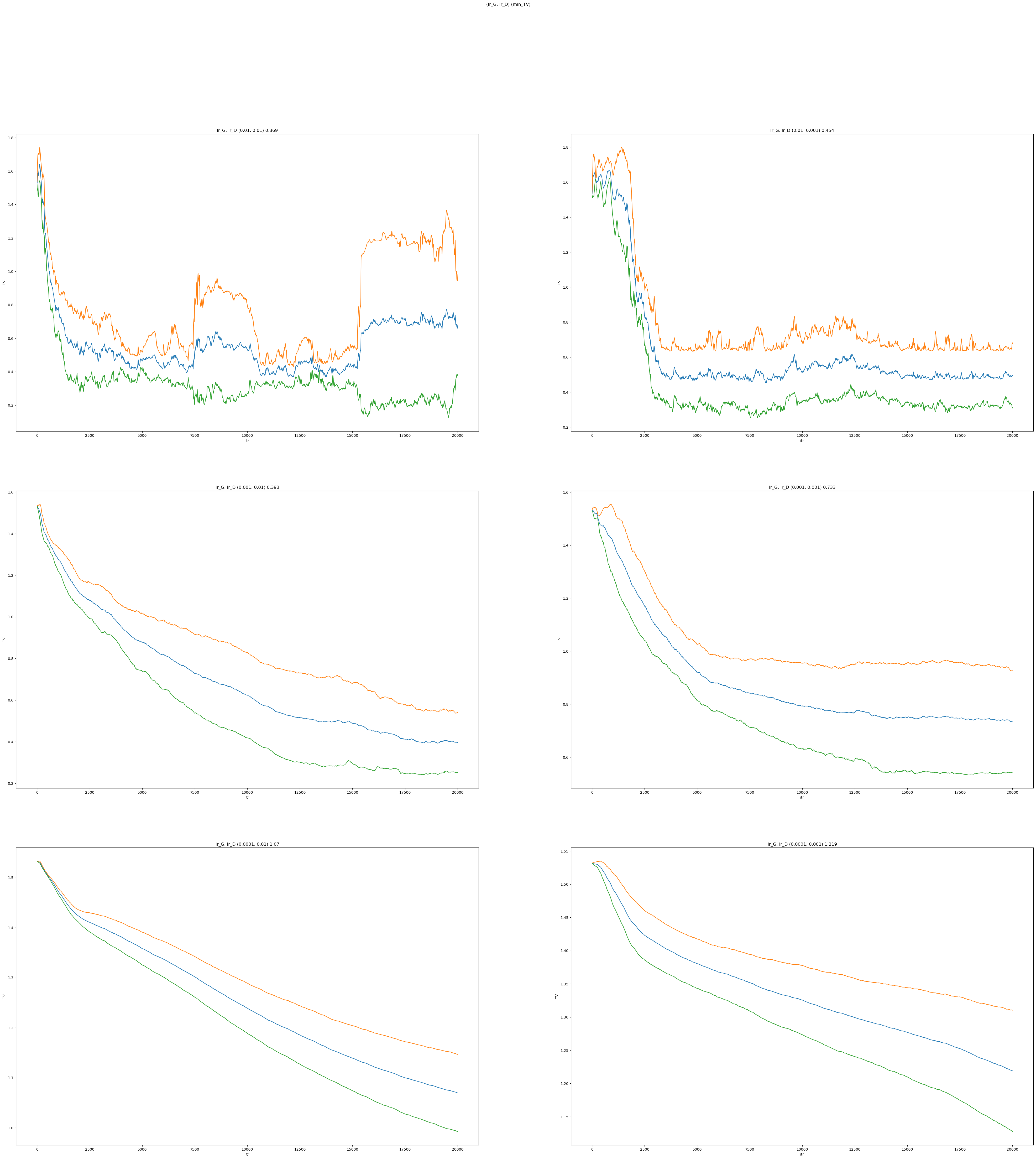}
    \caption{\textbf{QCBM+Inter-NS-MCR$^2$ with $\alpha=0.1$}: The total variation (TV) versus iteration number of the generated distribution learnt by the quantum circuit at various learning rates (lr) for the discriminator (D) and the generator (G). We vary G's lr at different rows ($0.01, 0.001, 0.0001$ top to bottom) and vary D's lr at different columns ($0.01, 0.001$ left to right). The blue line is the average value across $20$ runs, and the orange and green lines are one standard deviation above and below the average value.}
\end{figure*}

\begin{figure*}[h!]
\centering
    \includegraphics[scale=0.165]{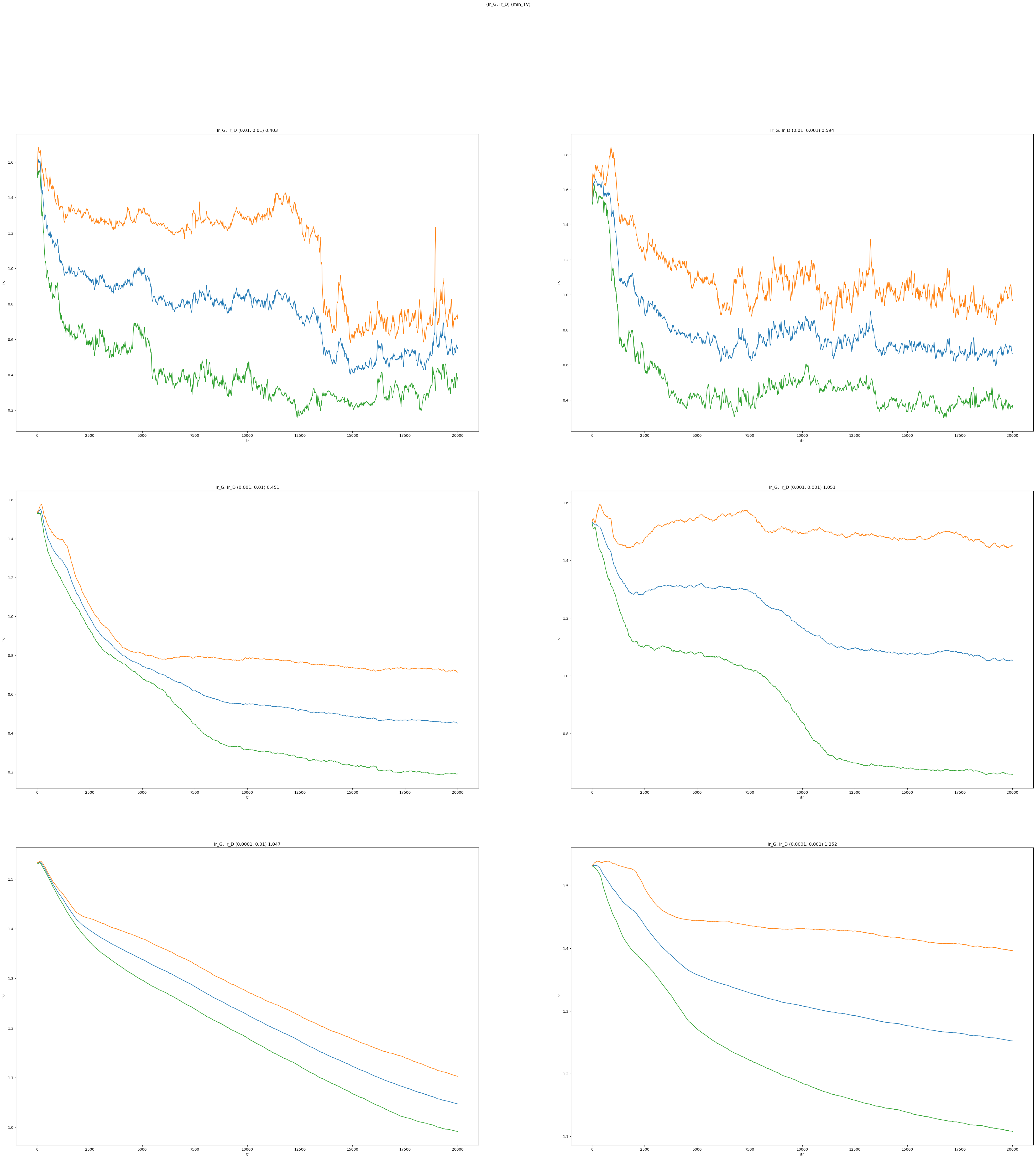}
    \caption{\textbf{QCBM+Inter-NS-MCR$^2$ with $\alpha=0.5$}: The total variation (TV) versus iteration number of the generated distribution learnt by the quantum circuit at various learning rates (lr) for the discriminator (D) and the generator (G). We vary G's lr at different rows ($0.01, 0.001, 0.0001$ top to bottom) and vary D's lr at different columns ($0.01, 0.001$ left to right). The blue line is the average value across $20$ runs, and the orange and green lines are one standard deviation above and below the average value.}
\end{figure*}

\begin{figure*}[h!]
\centering
    \includegraphics[scale=0.165]{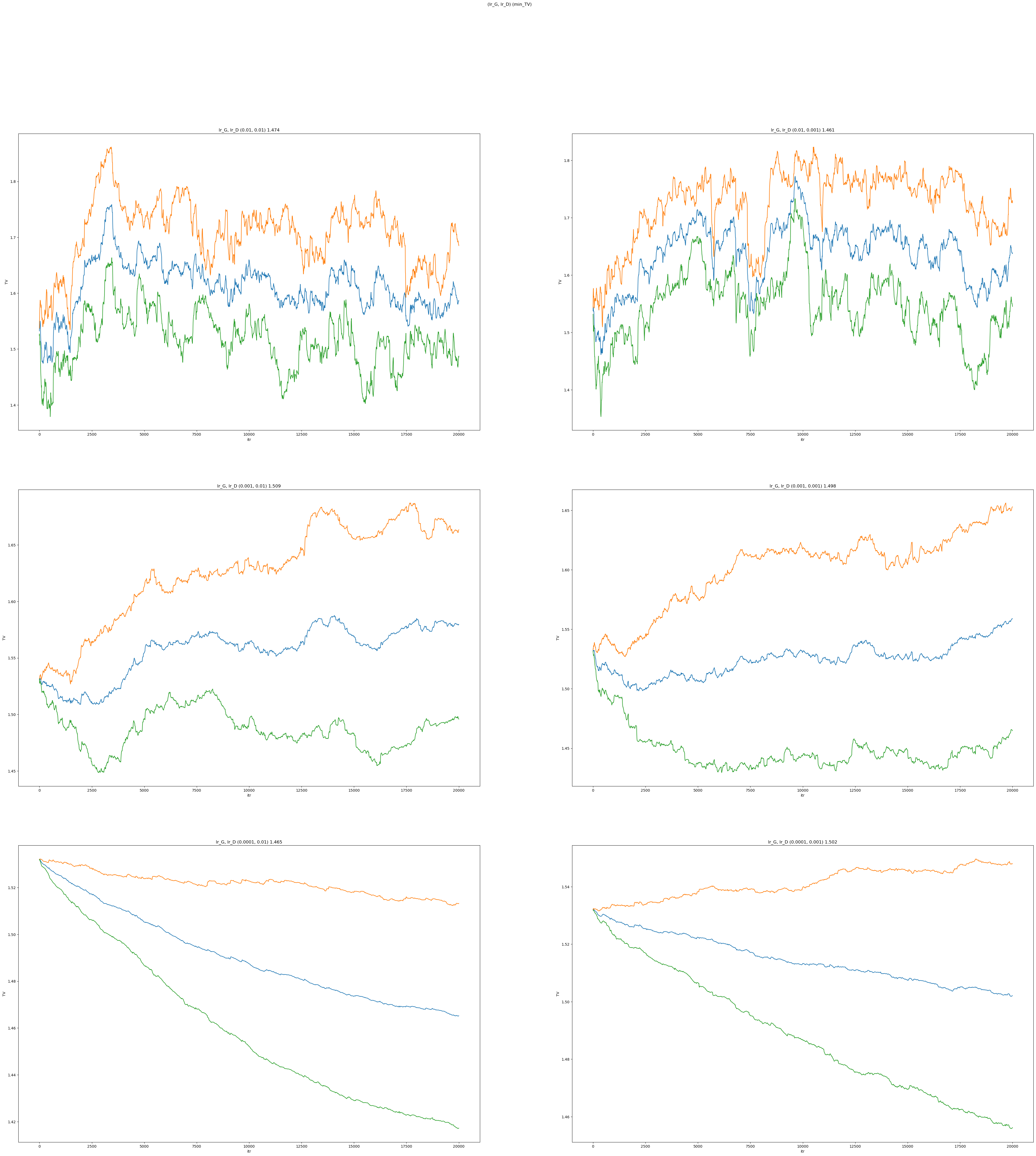}
    \caption{\textbf{QCBM+Inter-NS-MCR$^2$ with $\alpha=1.0$}: The total variation (TV) versus iteration number of the generated distribution learnt by the quantum circuit at various learning rates (lr) for the discriminator (D) and the generator (G). We vary G's lr at different rows ($0.01, 0.001, 0.0001$ top to bottom) and vary D's lr at different columns ($0.01, 0.001$ left to right). The blue line is the average value across $20$ runs, and the orange and green lines are one standard deviation above and below the average value.}
\end{figure*}

\begin{figure*}[h!]
\centering
    \includegraphics[scale=0.165]{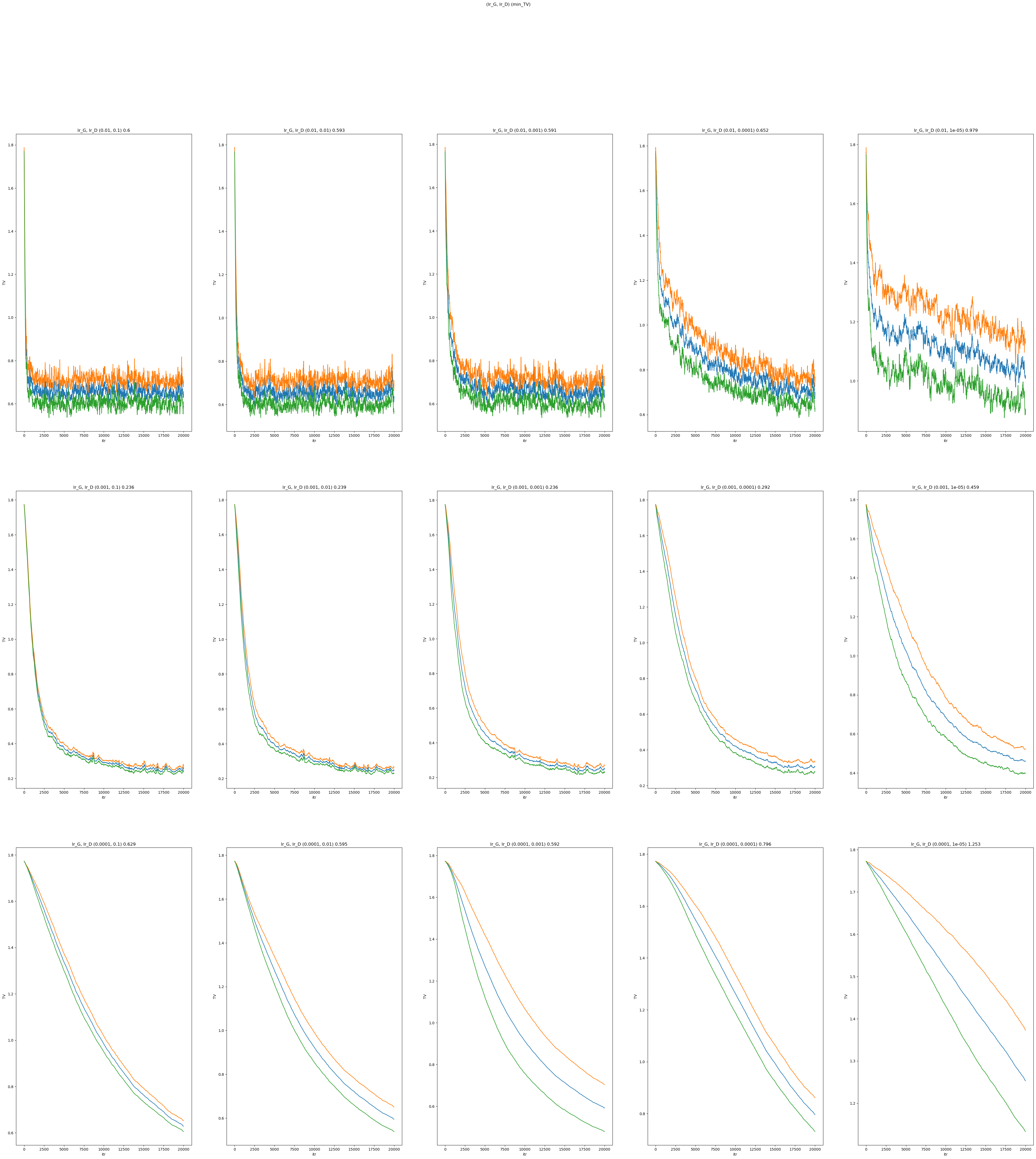}
    \caption{\textbf{QCBM+DNN-MMD}: The total variation (TV) versus iteration number of the generated distribution learnt by the quantum circuit at various learning rates (lr) for the discriminator (D) and the generator (G). We vary G's lr at different rows ($0.01, 0.001, 0.0001$ top to bottom) and vary D's lr at different columns ($0.1, 0.01, 0.001$, 1e-4, 1e-5 left to right). The blue line is the average value across $20$ runs, and the orange and green lines are one standard deviation above and below the average value.}
\end{figure*}

\end{document}